\documentclass[sigconf]{acmart}
\renewcommand\footnotetextcopyrightpermission[1]{} 

\usepackage{amsmath}
\usepackage{gensymb}
\usepackage[linesnumbered, ruled]{algorithm2e}
\usepackage{stfloats}

\usepackage{verbatim}

\usepackage[acronym,nohypertypes={acronym},shortcuts]{glossaries}
\usepackage{soul}

\usepackage{caption}
\usepackage{subcaption}

\usepackage{balance}

\begin{document}

\title[Your Car Tells Me Where You Drove]{Your Car Tells Me Where You Drove: A Novel \\ Path Inference Attack via CAN Bus and OBD-II Data}

\author{Tommaso~Bianchi}
\orcid{0000-0001-8192-5117}
\affiliation{%
  \institution{University of Padova}
  \city{Padua}
  \country{Italy}}
\email{tommaso.bianchi@phd.unipd.it}

\author{Alessandro~Brighente}
\orcid{0000-0001-6138-2995}
\affiliation{%
  \institution{University of Padova}
  \city{Padua}
  \country{Italy}}
\email{alessandro.brighente@unipd.it}

\author{Mauro Conti}
\orcid{0000-0002-3612-1934}
\affiliation{%
 \institution{University of Padova}
 \streetaddress{Via Trieste, 63}
 \city{Padua}
 \country{Italy}}
\affiliation{%
 \institution{Delft University of Technology}
 \streetaddress{Mekelweg 4, 2628 CD}
 \city{Delft}
 \country{Netherlands}}
\email{mauro.conti@unipd.it}

\author{Andrea Valori}
\affiliation{%
  \institution{Innova S.p.A.}
  \city{Trieste}
  \country{Italy}}
\email{andrea.valori@innovatrieste.it}

\begin{abstract}
Despite its well-known security issues, the Controller Area Network (CAN) is still the main technology for in-vehicle communications.  
Attackers posing as diagnostic services or accessing the CAN bus can threaten the drivers' location privacy to know the exact location at a certain point in time or to infer the visited areas.
This represents a serious threat to users' privacy, but also an advantage for police investigations to gather location-based evidence.
In this paper, we present On Path Diagnostic -
Intrusion \& Inference (OPD-II), a novel path inference attack leveraging a physical car model and a map matching algorithm to infer the path driven by a car based on CAN bus data. 
Differently from available attacks, our approach only requires the attacker to know the initial location and heading of the victim's car and is not limited by the availability of training data, road configurations, or the need to access other victim's devices (e.g., smartphones).
We implement our attack on a set of four different cars and a total number of 41 tracks in different road and traffic scenarios.
We achieve an average of 95\% accuracy on reconstructing the coordinates of the recorded path by leveraging a dynamic map-matching algorithm that outperforms the 75\% and 89\% accuracy values of other proposals while removing their set of assumptions. 

\end{abstract}

\keywords{Controller Area Network, On-Board Diagnostic, Path Inference, Location Privacy}

\maketitle
\pagestyle{plain} 

\newacronym{gps}{GPS}{Global Positioning System}
\newacronym{can}{CAN}{Controlled Area Network}
\newacronym{ids}{IDS}{Intrusion Detection Systems}
\newacronym{oem}{OEM}{Original equipment manufacturer}
\newacronym{obd}{OBD}{On-Board Diagnostic}
\newacronym{tpms}{TPMS}{Tyre Pressure Monitoring Systems}
\newacronym{osm}{OSM}{OpenStreetMap}
\newacronym{gpx}{GPX}{GPS eXchange Format}
\newacronym{ecu}{ECU}{Electronic Control Unit}
\newacronym{swa}{SWA}{Steering Wheel Angle}
\newacronym{sof}{SOF}{Start Of Frame}
\newacronym{rtr}{RTR}{Remote Transmission Request}
\newacronym{crc}{CRC}{Cyclic Redundancy Check}
\newacronym{ack}{ACK}{Acknowledgment}
\newacronym{eof}{EOF}{End Of Frame}
\newacronym{rke}{RKE}{Remote Keyless Entry}
\newacronym{icr}{ICR}{Instantaneous Center of Rotation}
\newacronym{opd}{OPD-II}{On Path Diagnostic - Intrusion \& Inference}

\section{Introduction}
Since the Jeep Cherokee remote hack by Charlie Miller and Chris Valasek~\cite{miller2015remote}, vehicle security has gained much more attention due to the consequences an attack can have on the security and safety of the people.
Indeed, the higher technological capabilities provided to modern cars by advanced sensors and communication technologies enlarge the attack surfaces of the vehicles. 
For this reason, vehicle security has recently played a significant role both in academia and industry. 
The most straightforward and diffuse attacks target the \ac{can}-Bus and its access to the vehicle due to its lack of security measures. 
Even if researchers proposed multiple defense mechanisms to safeguard access to the \ac{can} medium, e.g., intrusion detection systems and gateways~\cite{ids}, this environment's slow adaptation still threatens the vehicle's privacy and security.
Attackers can easily bypass the gateway connecting a malicious device
directly to the \ac{can} twisted paired cables.
An example of these attacks is illustrated by the \ac{can}-injection presented by the two automotive security researchers, Ian Tabor and Ken Tindell~\cite{canInjection}. 
They figured out a new attack used by thieves to impersonate a car key by message injection from the headlight.

Besides the security threat represented by the possibility of injecting safety-threatening messages in the \ac{can}-bus, vehicles pose a threat also to drivers' privacy.
Indeed, recent reports showed that vehicles collect and share a large amount of private data from their drivers \cite{mozilla1,mozilla2}.
Still, it is possible to gain sensitive data from cars nowadays.
Researchers already exploited various sensors to exfiltrate privacy information, such as the \ac{tpms}, brakes, or vehicle's speed. 
In the first case, a wireless device working on the same frequency of the \ac{tpms} (namely 433 MHz in Europe and 315 MHz in the rest of the world) can listen to the packet transmitted along the way by these sensors with a unique identifier and reconstruct the route of the vehicle~\cite{tpms_tracking}. 
Similarly, an attacker can locate a vehicle by identifying its keyless entry system and the key fob emanating the particular signals for vehicle-to-key communication~\cite{rke_tracking}.\\

\noindent{\textbf{Motivation.}} 
Among the personal information a malicious user could extract from a vehicle, location data represents one of the most sensitive information.
Indeed, a large body of literature explored the threats posed by location privacy and their relevance for profiling users in aspects such as home address, workplace, hobbies, sex, and age \cite{xiong2020adgan,qiu2020location,zhang2020decentralized}.
At the same time, gaining location information during a ride could be of great benefit for police investigations where officers need to gather evidence about illegal activities \cite{police1,koops2018location}.
Most vehicles nowadays mount a \ac{gps} sensor for navigation systems and location tracking in emergencies. 
Despite the straightforward connection between the \ac{gps} sensor and location data, it might not always be easy to gather such information.
Indeed, this sensor is not sending messages through the \ac{can}-Bus, unlike other sensors, and its data are hence not easily accessible by an external entity.
Also, in some situations, the \ac{gps} may not be able to reach the sensor on the vehicle, causing gaps in the location tracking.

Despite the availability of \ac{gps} sensors that the police enforcement, from now on referred as the attacker, could directly attach to the victim's car, we claim that such an approach would come with two major limitations.

First, the vehicle itself can work as a shield to the communication of such devices.
In this case, the attacker would not be able to gather precise location data and would infer a wrong path, hence limiting the effect of the attack.
Second, such an attack would simply be hindered by a victim using a \ac{gps} jammer.
These jammers are very cheap devices available to the public on online shopping websites, and they can block every signal in the \ac{gps} communication frequencies~\cite{jammers}. 
To develop an effective attack, we hence avoid assuming an attacker using externally connected devices.
As we show in this paper, focusing only on immediately available data once an external entity gains access to the \ac{can}-Bus, an attacker can perform precise vehicle path inference.
The attacker has the possibility to attach its device in hidden spots immediately after the \ac{obd} port. 
Furthermore, the actor can pose as a mechanic and attach devices on the \ac{obd} port to monitor the vehicle's diagnostic operations.
In a more sensible scenario, law enforcement could secretly access the car and connect the device in a few minutes.

Despite some works in the literature propose means to infer the path driven by a car \cite{dewri,gao,zhou,pese,sarker,waltereit}, they come with several limitations.
Indeed, they either achieve low accuracy on the inferred path or they do not provide generalization. In the first case, the lack of precise results is either due to very stringent assumptions about perfect alignment with road features (e.g., speed limit) \cite{dewri}, traffic patterns \cite{gao}, or the need to build a probabilistic model \cite{zhou}. 
In the second case, the lack of generalization is due to the need for the path inference algorithm of data from a specific area without the ability to process roundabouts and U-turns \cite{pese}, or due to the use of learning-based algorithms that do not provide deterministic decision \cite{sarker}.
The work achieving the best results is from Waltereit et al. \cite{waltereit}. 
However, the authors assume the availability of gyroscope and accelerometer data from a compromised smartphone.
Such an assumption limits the capabilities of the attacker, as exfiltrating data from a smartphone comes with significantly more challenges than exfiltrating data from a car.\\             

\noindent{\textbf{Contributions.}}
In this paper, we propose \ac{opd}, a novel path inference attack using easily accessible and reversible \ac{can} messages and \ac{obd} requests.
Our attack leverages a minimal and viable set of two assumptions: i) the attacker connects a collecting device directly to the victim's vehicle \ac{can} bus, and ii) the attacker knows the initial location, bearing (pointing or heading direction), and model of the victim's car.
The first assumption is common both in the literature~\cite{dongle} and has been proven viable in real-life car thefts~\cite{canInjection}.
The second assumption includes information that is easily accessible by anyone and does not present any challenge for the attacker.
The central intuition behind our attack is that we can reconstruct the path driven by a vehicle leveraging the Bicycle physical model~\cite{bicycle} and a map-matching algorithm.
Given the vehicle's starting point and bearing, we leverage the bicycle model to calculate the vehicle's covered distance during a given time window.
Thanks to this model, we build a point-to-point \ac{gpx} file with the inferred coordinates.
We cross-reference the inferred traveled path with a map-matching algorithm to obtain a precise match with existing roads.
To this aim, we leverage the Valhalla~\cite{valhalla} open-source project together with a collaborative map, \ac{osm}~\cite{osm}, also freely available.
Applying a map-matching algorithm reduces the computation error, re-setting the vehicle coordinates to the correct path at each time window. 

Such an approach leads to a deterministic inferred path without the need to compute a group of possible tracks to be tested to infer the correct one. 
Since we do not rely on any data-based training process, road conditions, or traffic pattern, our approach provides high adaptability without the need to re-train a model or to overcome the limitations imposed by road configurations (e.g., roundabouts).
The average efficacy of our algorithm is 95\%, with peaks of perfect reconstruction. 
The deterministic outcome with a high confidence level outperforms the state-of-the-art algorithms in the literature. 
The following are the main contributions of our work:
\begin{itemize}
    \item We propose \ac{opd}, a novel deterministic path inference attack. Our approach leverages \ac{can} and \ac{obd} messages to reconstruct the path driven by a victim vehicle during a certain time window and cross-reference it to real-world maps via a map-matching algorithm.
    \item We developed our attack on a set of four real cars with a total number of 41 paths. We provide a thorough explanation of the steps involved in inferring the required data and the setting of parameters. Our results show that we can reconstruct the traveled path with an average 95\% efficacy.
    \item We release our code as open source to facilitate the reproduction of our results and foster future studies on the security of path inference via \ac{can} bus messages.

\end{itemize}

The paper is structured as follows: in Section~\ref{sec:background} we discuss the background and related works the reader needs to understand this manuscript. In Section~\ref{sec:model}, we present the threat model, while we show the steps of the attack in detail in Section~\ref{sec:attack} with our implementation in Section~\ref{sec:implementation}. In Section~\ref{sec:evaluation}, we provide the analysis and results of the presented algorithm for the attack. Finally, we discuss the contribution, the drawbacks, and future works in Section~\ref{sec:discussion}, and the conclusions in Section~\ref{sec:conclusion}.

\section{Background and Related Work}\label{sec:background}
In this section, we discuss the background (Section \ref{subsec:background}) and the literature review of the state of the art (Section \ref{subsec:related}). 

\subsection{Background}\label{subsec:background}
The path inference attack is against a driver's privacy, exploiting the simplicity of data retrieval in a vehicle. 
As pointed out in the Introduction Section, vehicles suffer from different attack types. 
In particular, injection is still a threat from \ac{can} despite many researchers presenting works to defend against such intrusions.
In Section~\ref{subsubsec:can}, we introduce the (in)famous \ac{can}-Bus and the diagnostic protocol working over it, the \ac{obd}, that we leveraged to perform our location privacy attack. Finally, Section~\ref{subsubsec:gps} discusses the technology and privacy concerns of the \ac{gps} technology.

\subsubsection{CAN Bus and OBD-II}\label{subsubsec:can}
The \ac{can}-Bus is a standard defining the physical and data link layers (ISO 11898~\cite{iso11898}) for robust in-vehicle communication networks between \acp{ecu}.
It is extremely simple and robust, with the advantage of a very low cost.
There are multiple \acp{ecu}, each controlling and providing information about different sensors and actuators to run the vehicle.
Two twisted wires form the \ac{can} High and \ac{can} Low and connect \acp{ecu}. 
The dominant bit is 0, while the recessive bit is 1. 
The dominant bit is used to determine who can talk over the medium.
The communication happens without encryption or message authentication; therefore, forging messages is trivial.
Furthermore, the only mitigation available in cars is a gateway between the \ac{obd} access port and the \ac{can}-Bus. 
Still, a malicious person can bypass it easily by connecting directly to the \ac{can} from another access point, such as the cables, or interposing its instruments after the gateway.
In the scope of this work, we are interested in the \ac{can} frame that is divided into eight fields:

\begin{description}
    \item[\ac{sof}] It is a dominant 0 bit to tell that the node intends to talk on the channel.
    \item[ID] This is the frame identifier and indicates the priority on the Bus. Lower values have higher priority. It represents the importance of the content, and even if it is not coupled with an ID for the \ac{ecu}, this is usually the case in practice. However, the producers tend to assign their own IDs, and the reversing task is still an open problem.
    \item[\ac{rtr}] It distinguishes whether a node sends or requests data.
    \item[Control] It contains the information about the extended identifier, if present, and the four bits to specify the data field length. 
    \item[Data] This is the payload of the frame. The content is encoded in a format that depends on the data type and the manufacturer-specific implementation. Also, the reverse engineering task is nonetheless an open research concern.
    \item[\ac{crc}] This field ensures data integrity.
    \item[\ac{ack}] The slot marks if the node has received the data correctly.
    \item[\ac{eof}] End of the frame.
\end{description}

\begin{figure}[!h]
    \begin{center}
    \includegraphics[width=\columnwidth]{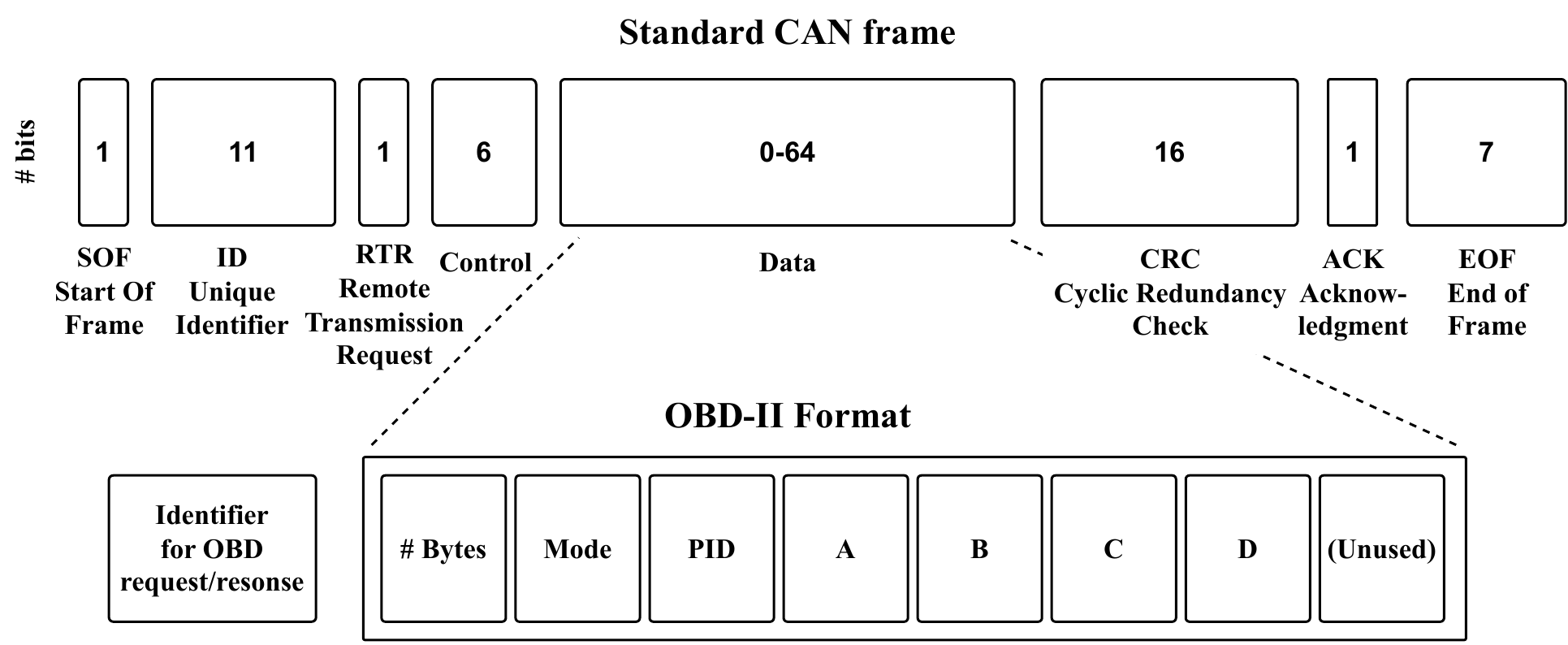}
    \end{center}
    \caption{\label{fig:canobd} The CAN frame uses the Identifier to perform medium contention. The data field contains information to use during diagnostic operations with the OBD protocol.}
\end{figure}

The most important fields for this attack are the ID and Data fields. We inject and log packets with specific values to get the IDs for the \ac{swa} sensor and the vehicle's speed. 
We obtain this last value by exploiting the \ac{obd} requests and responses standardized across different vehicles and vendors. 
Figure~\ref{fig:canobd} represents the packet structure in \ac{can}-Bus and the diagnostic packet working over it. 

\begin{figure}
    \begin{center}
    \includegraphics[width=\columnwidth]{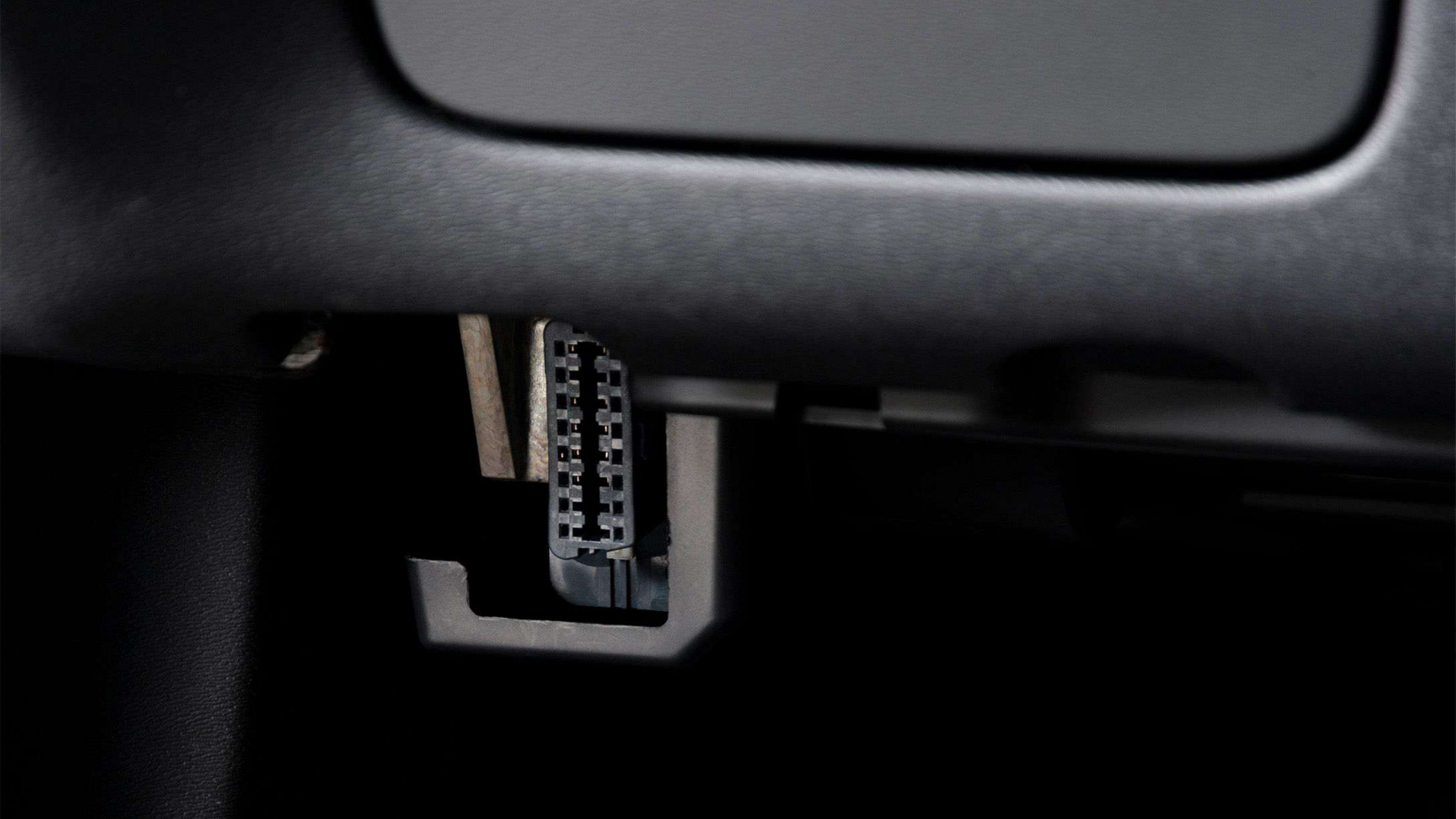}
    \end{center}
    \caption{\label{fig:obdport} Example of OBD port connector on board a vehicle. It is usually located under the steering wheel, and it has easy access for diagnostic operations.}
\end{figure}

Over the \ac{can}-Bus, different higher layer protocols can run to transmit various information.
Among them, \ac{obd}-II transports diagnostic data valuable for understanding and resolving problems inside the vehicle and the \ac{ecu}s. 
Mechanics and car owners (but also attackers) can easily access it through the \ac{obd} port.
We report an example of the location and the port in Figure~\ref{fig:obdport}. 
It works sending protocol fields inside the data of a \ac{can}-Bus frame and utilizing specific IDs, as we describe in the following:
\begin{description}
    \item[\ac{obd} Identifier] It uses the standard 11-bit identifier, with a fixed ID value of \textit{7DF} for request messages, and IDs from \textit{7E8} to \textit{7EF} for response messages.
    \item[Length] Number of bytes of the \ac{obd} message. 
    \item[Mode] It indicates the mode and type of request/response.
    \item[PID] Each mode has its standard identification for the values and encoding formula.
    \item[Values and Unused] The actual data in bytes. The last byte is not used in this protocol.
\end{description}

\subsubsection{Location Privacy}\label{subsubsec:gps}
The capacity to track a vehicle without having access to a \ac{gps} sensor is of particular interest for public agencies and intelligence that aim to discover possible hidden spots and locations of criminals and people under surveillance. 
Criminals know about the possibility of being tracked with a \ac{gps} and can try to evade it with probes to identify possible bugs inside the car.
Moreover, lawbreakers can use jammers to block every \ac{gps} communication and evade location tracking by law enforcement.
Furthermore, sometimes it is hard to get the signal outside the hidden location of a \ac{gps} sensor that could notice the data transmission shielded by the vehicle's metal body. 
In these cases, exfiltrating location data without errors or being caught is essential for public security reasons and investigations.

\begin{table*}[!b]
    \centering
    \caption{Related Work Table}
    \resizebox{1.8\columnwidth}!{
    \begin{tabular}{p{1.8cm}|p{2cm}|p{1.7cm}|p{1.9cm}|p{1.7cm}|p{1.7cm}|p{2.1cm}|p{2cm}}
    \hline
         & \textbf{Dewri~\cite{dewri}} & \textbf{Gao~\cite{gao}} & \textbf{Zhou~\cite{zhou}} & \textbf{Pese~\cite{pese}} & \textbf{Sarker~\cite{sarker}} & \textbf{Waltereit~\cite{waltereit}} & \textbf{\ac{opd} (Our proposal)} \\
    \hline
       \textbf{Data Source} & Speed from Tracking Device & Speed from OBD & Speed from OBD & Data Collection Platform & SWA & OBD and Smartphone &\textbf{ Speed and SWA from OBD} \\ \hline
       \textbf{Reference Data} & Distance Travelled and Timestamp & Maps & Maps & Maps & Maps and real-time traffic & Maps, smartphone, and real-time traffic & \textbf{Maps} \\ \hline
       \textbf{Method}  & DFS & Elastic Patching & Hidden Markov Model and DFS & Road Curvature Matching & Brake-Based Location Tracking & Distance and direction route inference & \textbf{Dynamic Map-Matching} \\ \hline
       \textbf{Starting Point Needed?}  & no & no & no & yes & no & yes & \textbf{yes} \\ \hline
       \textbf{Accuracy}  & 37\% & 14\% - 250 m error & 70\% in Top 30 & 71\% & 89\% in Top 5 & 78\% with 100\% accuracy, 96.5\% with 75\% accuracy & \textbf{95\%} \\
    \hline
    \end{tabular}}
    \label{tab:related}
\end{table*}

\subsection{Related Work}\label{subsec:related}

Only a few works in the literature deal with path inference from \ac{can}-Bus data. 
The different approaches significantly improved the accuracy of the inference, starting from a speed-based technique to a more advanced machine learning and brake-signal-based approach. 
All the related work consider the possibility of reconstructing the paths by an insurance company with access to some data considered harmless, such as speed and steering wheel angle.

Dewri et al.~\cite{dewri} proposed a first location inference attack using simple driving features. 
They tried to extrapolate turns and stops from the data collected and, knowing the starting point, to generate the shortest path to the endpoint. 
They used the shortest path because the authors did not want to guess the correct intermediate path between the starting and ending nodes.
The proposed ranking algorithm could deduce correctly around 37\% of the paths. 
The primary limitations regard the need for perfect alignment with speed limitations, the actual movement speed of the vehicle, and the absence of traffic to conduct the inference.

A simpler approach is reconstructing the path using only the speed data and knowing the home location. This algorithm is called \textit{Elastic Patching}~\cite{gao}.
They used \ac{osm} to identify routes given the vehicle's speed, considering the streets free of obstacles and traffic.
The speed depends on the type of action and curvature of the road, allowing the attacker to understand the path with the stretching and compression of the speed profile.
There are two significant drawbacks to this approach.
First, it cannot work in normal traffic conditions, and second, it needs to assume the roads are distinguishable. 
The total accuracy is 14\%, considering an error of less than 250 meters. 
Even if the results did not have an enormous impact, this work showed how a single type of data can harm the location privacy of a driver.

In 2017, researchers developed a new speed-based trajectory inference algorithm~\cite{zhou}.
Like in~\cite{gao}, the authors exploit public information about the roads, such as speed limits, to build the possible street crossed by a vehicle. 
To overcome the limitations of \cite{gao}, the researchers added the real-time traffic information and used the Hidden Markov model to consider the probability of taking a certain way, gaining a 70\% probability of having the correct path in the top 30 inferred routes.

Pesè et al.~\cite{pese} and Sarker et al.~\cite{sarker} shifted towards the use of other sensor data to locate a vehicle, using the \ac{swa} and brake signals, respectively. 
The first paper poses as an insurance company that wants to gather the road crossed by a vehicle with data shared by the users. 
In their work, the authors pre-process the data of a specific area where the matching algorithm can work based on the curvature of the streets, achieving a total of 71\% to retrieve the correct route. 
Consequently, the algorithm needs a re-train for a new city, and the conformation of the roads limits the attack.
In fact, it cannot work on maps similar to a grid, such as Manhattan, as the authors report.
Additionally, the algorithm doesn't consider roundabouts and U-turns, limiting the efficiency of real-world scenarios.
Instead, \cite{sarker} proposes a classifier to infer the action taken by a vehicle through the brake signals. 
The authors combine this brake information with real-time traffic and road characteristics to identify the crossed path based on the maneuvers. 
In this case as well, the algorithm cannot work well in the presence of very close streets during a turning maneuver.
The manuscript reports an accuracy of around 89\% to retrieve the correct path, but unfortunately, the setting description is not precise.
The algorithm returns the best \textit{K} paths, and the ending point is retrieved clustering predicted destinations, restricting it to a particular area.

Similar to our work~\cite{waltereit}, in 2019, a new algorithm presented by Waltereit et al. can infer the path in a specific region without using starting and destination points from speed and direction.
It is able to derive the traveled route 78\% of the time if considering a perfect score in route matching. 
If the track comparison bound is set to 75\% in accuracy (lower precision), the success rate is 95.5\%.
The authors assume that a gyroscope or accelerometer is available via a smartphone to gather the turning actions. 
This assumption increases the difficulty and practicability of the attack in a real-world scenario.

Unfortunately, none of these works provide code and means to reproduce the results. In Table~\ref{tab:related}, we show the differences between the different works.

\section{Threat Model}\label{sec:model}
\begin{figure*}[!h]
    \begin{center}
    \includegraphics[width=0.6\textwidth]{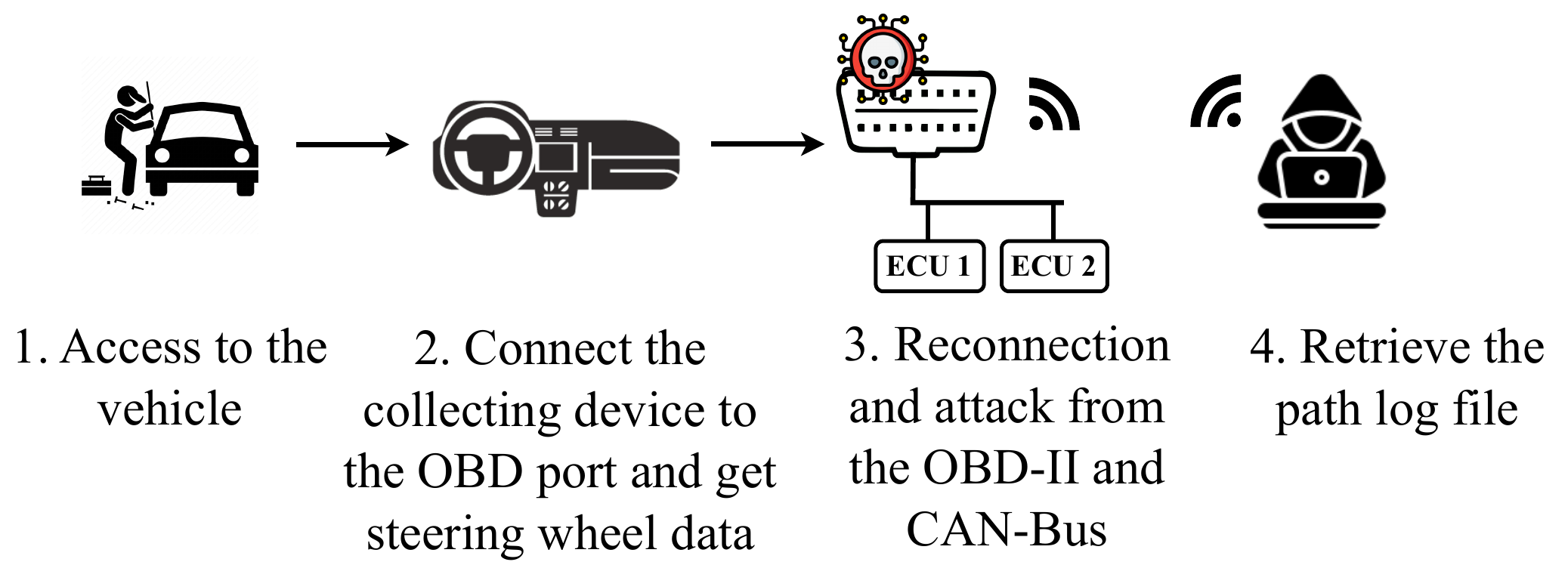}
    \end{center}
    \caption{\label{fig:scenario} The threat model requires entering the vehicle and installing the malicious device to connect to the CAN Bus and exfiltrate data.}
\end{figure*}

In this section, we describe the attacker's capabilities, assumptions on the model, and a general data collection procedure to use for the attack. Figure \ref{fig:scenario} depicts the attack process. \\

\noindent{\textbf{Attacker's capabilities.}}
We consider an attacker that can enter the target vehicle. 
In most vehicles available on the streets, it is trivial to force the way in, as proven in recent news, such as stealing a car using \ac{can}-Bus injection through the vehicle's headlight~\cite{canInjection}.
Kits to break into vehicles are available online in the most famous markets like Amazon, naturally with a disguise for the real purpose. The kit contains a long-reach grabber, air wedge pump, and other tools for car lock-picking.
Once inside the vehicle, the attacker can collect data from the \ac{obd} port by connecting it to a malicious data collection and communication device.
We henceforth refer to this device as the \textit{collector}.
Alternatively, they can tamper the \ac{obd} port to bypass the gateway and insert the \textit{collector} after the security module.
The connection to the \ac{obd}, or even better to the twisted pair cables, gives the attacker direct access to the \ac{can}-Bus and all its traffic.
In this way, the attacker can inject and eavesdrop on the packets of interest to extrapolate data about the \ac{swa} sensor, reverse engineer the sensor's \ac{can} ID, and know the instantaneous vehicle's speed through standard \ac{obd} request/response.  \\

\noindent{\textbf{Assumptions.}}
To deliver our novel attack, the attacker requires access to the vehicle. 
Furthermore, either the car does not present a \ac{gps} sensor, or reading its data is not allowed or impossible for the malicious actor due to the dedicated wires and communication channel for this sensor inside the vehicle. 
A dedicated wired connection would be hard to access in the opposite of the \ac{obd} port.
Moreover, the application of a \ac{gps} bug suffers from the difficulty of receiving and sending signals outside the metal chassis of a vehicle.
We also consider that GPS bug detectors are easily available online, and the victim might employ such a device to easily detect if they are tracked.
The attacker can enter the vehicle and tinker with the \ac{obd} port. 
When entering the victim's car, the attacker takes note of the vehicle's starting point in the form of latitude, longitude, and heading direction.
This information is essential for the next steps of the attack. \\

\noindent{\textbf{Data Collection.}}
The attacker connects the \textit{collector} to the \ac{can}-Bus medium channel to collect and store driving data.
The \textit{collector} is equipped with communication capabilities so the attacker can easily instruct it to deliver the attack or retrieve the collected data.
A low-resource device, such as a Raspberry Pi, with \ac{can}-Bus and wireless communication capabilities is sufficient to carry out the attack. 
Through the wireless channel, the malicious actor can access the device via protocols like SSH and start the scripts for the attack.
While the \textit{collector} can constantly monitor the \ac{can}-bus data, we assume the attacker has limited interaction times with it.
Such interactions are limited to the periods when the victim's car is parked or close to the attacker.
The attack requires a minimum number of two interactions between the attacker and the \textit{collector}.
Before the victim's departure, the attacker must interact with the \textit{collector} to start the data collection process.
At the end of the victim's trip, the attacker connects again to the device, stops the attack, and retrieves the packet logs to perform the path inference step offline.
The data retrieval procedure requires the attacker to be within the device's wireless network range.
Equipping the device with a cellular network module and communicating through the internet is also possible.

Concerning the victim's capability to spot the attack, the wireless transmission channel (e.g., WiFi or LTE) is more challenging to detect or block due to the variety of devices working on the same frequencies.
Additionally, the target person should block with a jammer all the communications of its own devices, e.g., the mobile phone.

\section{Attack and Technical Approach}\label{sec:attack}
In this section, we describe every step of the attack. 
The attack comprises six steps, starting from the reverse engineering of the steering angle and data collection discussed in Section~\ref{subsec:rewheel} and Section~\ref{subsec:obdrop}, respectively. 
We provide details of the path inference methodology in Section~\ref{subsec:opd}. 

\subsection{Physical Access}\label{subsec:access}
The attacker's first step is gaining access to the vehicle, attaching the device we described in Section~\ref{sec:model}, and performing a first packet logging and evaluation.
Access can be achieved in different ways.
The intrusion can be mechanical, similar to old thefts, or exploit much more advanced techniques such as \ac{rke} systems relay attacks~\cite{relay}.
In the scope of this work, we assume access to the vehicle is possible, and that the hostile user connects the collecting device to the \ac{obd} port.
Practical examples are the masquerading of an attacker as a mechanic or car service personnel. The malicious actor can access the vehicle and attach a device collecting the data without the owner noticing it. 
Another example is the collection of \ac{can}-Bus data from a truck fleet that is expected to be privacy-preserving, but we show that speed and steering wheel angle are enough to infer the path of a car. 

\subsection{Reverse engineering the Steering Wheel Angle Sensor}\label{subsec:rewheel}
Once inside the vehicle and with the \textit{collector} attached to the \ac{obd}, the attacker must identify packets that encode the car's direction. 
To this aim, they perform a preliminary reverse engineering process to retrieve the \ac{can} ID of the \ac{swa}.

The encoding for the \ac{swa} sensor usually involves two bytes in the data field of the \ac{can} packet.
There is no standard for the encoding and position of the bytes, but we observed that the format is consistent across different vendors.
The idea behind the reverse engineering approach is that minimal variations in the \ac{swa} cause minimal variations in the data field of \ac{can} bus packets encoding such information.
The attacker moves the steering wheel left and right with the vehicle in a steady state while logging the \ac{can}-Bus packets.
In this way, the data field continuously encodes the wheel's angle shifts at regular time intervals with minimal variations.
In the perfect straight direction of the wheels, the packet contains the offset value (e.g., \textit{0x0000} or \textit{0x7FFF}).
Turning right decreases the field's content, while the opposite, turning left, increases the value with respect to the offset.
For example, consider the offset \textit{0x7FFF}, with values over \textit{0x8000} for turning left and values lower than the offset as turning right. 
Considering it as angles, the decoded values can range from -50 (completely right) to +50 (completely left).
Using the \textit{hamming distance} between successive data filed values can make the ID identification process easier.
To recall, the hamming distance between two vectors in information theory is defined as the number of positions at which their corresponding symbols are different.
The attacker can use the average hamming distance computed between every two consecutive packets for each ID to filter packets that never change in the captured log file.
Additionally, the \ac{swa} sensor has a high priority in the \ac{can}-Bus, allowing the attacker to filter out also for very high IDs.

At this point, the attacker needs to reverse-engineer the decoding formulation of the steering angle sensor.
This is the most manual phase of the attack, which aims to find the offset and decoded values for the sensor.
The attacker has to perform some checks manually to get the correct decoding. 
Generally, the decoding conversion is similar across different vehicles, especially if the car manufacturer produces car components for various brands.
Even if not standardized and empirically determined through reverse engineering processes, the simpler decoding formula is as follows:
\begin{equation}\label{eq_angle}
    angle = ((b_0 \times 256 + b_1) - offset) \times c,
\end{equation}
\noindent
where $b_0$ and $b_1$ are the two bytes corresponding to the sensor data, and the offset is the corresponding centered values, like in the previous example. 
The decoded angle often results in over-scaled values, hence the correction by a constant value \textit{c}
Sometimes, the decoding can be more complex than this, adding the two's complement computations to get the negative values. 
The steering wheel logging is the only part of this phase that needs to be performed online. 
Once the reverse engineering phase is over, the attacker knows of the \ac{swa} ID and the angle decoding.
After the first logging phase while moving the steering wheel, the rest of the reverse engineering part is performed offline.
    
\subsection{Data Collection}\label{subsec:obdrop}
Whenever the attacker is in the range of a previously installed \textit{collector}, it can start the logging phase.
Leveraging the \ac{swa} ID identified in the previous step and using the \ac{obd} protocol standard packets, the attacker can filter out the unwanted packets and collect only the ones of interest for path inference.
Concerning the \ac{obd} packets, it is easy to filter them because they are standardized across different vendors.
In fact, the ID used to give \ac{obd} response is\textit{0x7E8}.
In this attack, the knowledge of the speed is carried in the \ac{obd} packet of this type. 
Consequently, it is trivial to retrieve it.
To request the instantaneous speed values, the \ac{obd} standard defines the following message:
\begin{verbatim}
7DF#02010DAAAAAAAAAA
\end{verbatim}
\noindent
The part before the \textit{\#} symbol specifies the ID. 
The aforementioned $0x7DF$ ID is the standard request \ac{can} ID for the speed.
The part after the hash symbol contains the request parameters, where the different fields represent:
\begin{description}
\item[Data length] It is the number of bytes of the request, excluding the length byte. We have $0x02$ in this request, specifying a length of 2 bytes.
\item[Mode] $0x01$ This mode indicates to take the freeze frame data.
\item[PID] $0x0D$ The PID describes the identifier for the vehicle's speed.
\end{description}
\noindent
The rest of the data field is filled with As and unused by the diagnostic protocol.

During the data collection, the attacker waits for the vehicle to return while the on-board malicious device logs the packets along the trip.

\subsection{Data Retrieval}\label{subsec:retrieval}
At the end of the journey, the attacker can reconnect to the device and stop the logging, pull the log file from the device, and run the dynamic map-matching algorithm to infer the route followed by the vehicle.
The rest of the procedure is offline, and access to the vehicle is not required.

\subsection{Deterministic Path Inference}\label{subsec:opd}
This is the core step of the attack.
In this phase, the attacker inspects the retrieved packets to create the \ac{gpx} points of the track.
The overall attack needs some data retrieved during the previous phases, and they are as follows:
\begin{itemize}
    \item Car model or manufacturer: this specifies the correct \ac{swa} ID and angle decoding to use. The attack can get them from previous phases.
    \item Vehicle's wheelbase: the horizontal distance between the center of the front and the rear of the wheels. This is usually a standard measure for all vehicles of the same model and make.
    \item Starting point: latitude and longitude of the vehicle at the starting point of the attack. The attacker can get them with any positioning system.
    \item Heading (bearing): it is the car's direction in a range starting from 0 to 360. The 0 value points to the North. The bearing can be noted down during the first attack step with a compass or phone sensor. This information can be easily found on the internet or in the vehicle's manufacturer.
\end{itemize}
These data are easily retrievable for the attacker because it needs to access the vehicle, and hence, it knows the \textit{(longitude, latitude)} coordinates and the bearing.
Additionally, the attacker can define the maximum physical steering angle possible for the vehicle. 
This is $\pm35$ degrees (left and right) by default due to the physical limitations of cars.
These values are essential for the physical model to work correctly.
Algorithm~\ref{alg:inference} provides the steps of the path inference performed via the dynamic map-matching.
\begin{algorithm}[!h]
\caption{Path inference starting from the \ac{can}-Bus log file.}\label{alg:inference}
\footnotesize
\KwData{$Logfile, (Vehicle, Wheelbase), (Lat, Long, Bearing)$}
\KwResult{$Inferred\; Path\;  as\;  GPX\; coordinates$}
$counter_{interpolation} \gets 0$\;
$list_{coordinates} \gets empty$\;
$list_{inferred} \gets empty$\;
\While{$len(total\; packets) \geq 0$}{
    $Angles, Speeds \gets packets\; in\; t_{window}$\;
    $AVG_{angle} \gets avg(Angles)$\;
    $AVG_{speed} \gets avg(Speeds)$\;
    $Distance \gets AVG_{speed} \times t_{window}$\;
    \If{$Remaining\; Distance > 0$}{
        \tcc{Adjust distance}
        $Distance \gets Distance + Remaining\; Distance $ 
    }
    $Adjust\; AVG_{angle}\; with\; steer_{max}$\;
    \tcc{Apply bicycle model for new heading direction}
    $Bearing \gets Bicycle\_Model(AVG_{angle}, Wheelbase, t_{window})$ \;
    $Adjust\; bearing\; based\; on\; maximum\; speed\; speed_{max}\; for\; turning$\;
    $Get\; bearing\; on\; 0\; to\; 360\; value\; interval$\;
    \tcc{Compute new point's coordinates}
    $(Lat, Long) \gets geodesic\_distance(Lat, Long, Bearing, Distance)$ \;
    \tcc{Save point in the list for interpolation}
    $list_{coordinates} \gets list_{coordinates} + (Lat, Long)$ \;
    \If{$counter_{interpolation} \% max\_interpolation\_points = 30$}{
        \tcc{Apply Map Matching algorithm}
        $coordinates_{inferred} \gets Map\_Matching(list_{coordinates})$ \;
        \tcc{Save inferred coordinates from Map Matching}
        $list_{inferred} \gets coordinates_{inferred}$ \;
        \tcc{Get possible remaining distance}
        $Remaining\; Distance \gets abs(compute\_distance(coordinates_{inferred}), list_{coordinates})$ \;
        $list_{coordinates} \gets empty$\;
    }
    $counter_{interpolation} \gets counter_{interpolation} + 1$\; 
} 
\end{algorithm}

Once the algorithm receives as input the information about the initial point and the car model, it computes the average of the speed and angle values from the log file over a time window $t_{window}$.
Each packet contains information about the \ac{swa} sensor and the speed, retrieved using the decoding reversed during the reverse engineering step discussed in Section~\ref{subsec:rewheel}.
Speed is converted in meters per second to allow further computation in the International System of Units.
From the speed, the algorithm computes the distance the vehicle covers in a given time frame.

The inference uses an interpolation of $max\_intepolation\_points$ time windows (namely, a time frame of $n$ windows of time $t_{window}$) to avoid keeping the vehicle in the same position due to the small step taken with the use of a smaller time window.
In fact, having a short window could place the car in the same \ac{gps} coordinates instead of moving it further in the track during the inference using map-matching.
In the time step $t_{window}$, the algorithm computes the vehicle's actual movement, adopting the simplest available physical model: \textit{the bicycle model}~\cite{bicycle}.
Essentially, the model approximates the four-wheel vehicle as a two-wheel one, simplifying the computation of a rigid corp movement.
Furthermore, we are only interested in the dynamics to bring the vehicle in the new heading direction after the maneuver.
Before applying the model, the average angle is adjusted between the maximum thresholds, namely $steer_{max}$, in positive or negative depending by the turn direction.
\begin{figure*}[!b]
    \begin{center}
    \includegraphics[width=\textwidth]{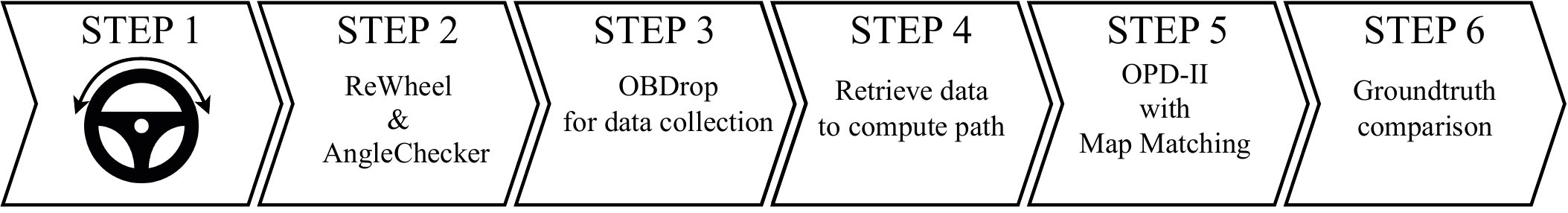}
    \end{center}
    \caption{\label{fig:workflow} The six attack steps once the device is connected. The attacker needs to reverse the angle sensor, and two tools are available for this purpose in Step 2. After that, the attacker collects data and exfiltrates it to use the OPD-II tool for the dynamic map-matching algorithm and infer the path.}
\end{figure*}
The physical formulation for the new bearing starts with the computation of the angular speed $\omega$ as 
\begin{equation}\label{eq_angle}
\omega = speed \times tan(radians(angle)) / wheelbase,
\end{equation}
\noindent
where the speed is the average speed in the time window, multiplied by the tangent trigonometric function applied to the angle's radians value of the average angle ($AVG_{angle}$ in Algorithm \ref{alg:inference}) for the time step $t_{window}$.
Then, the actual bearing respects the North direction (0 in the bearing's values range) as
\begin{equation}\label{eq:new_angle}
        bearing = atan2(sin(\omega \times t_{window}), cos(\omega \times t_{window})),
\end{equation}
\noindent
where $t_{window}$ is the time window adopted, $sin$ and $cos$ are the sine and cosine trigonometric functions, and $atan2$ is a variation of the arctangent trigonometric function.
In Appendix~\ref{appendix:bicycle}, we describe the physical model in more detail.
Suppose the vehicle's speed is over a maximum speed $speed_{max}$ allowed for turning. 
In that case, we also adjust the heading direction using the bearing between the starting and ending point of the previous time window, considering that a turn is impossible.
In this way, the car will be in the exact straight direction.
The computed bearing is subtracted from the current heading direction, and we perform an adjustment to keep the value in the correct interval from 0 to 360.
On top of that, after collecting the \textit{n} ($max\_interpolaton\_points$) time windows, the algorithm adjusts the distance based on the previous interpolation.
The algorithm computes the length depending on the difference calculated between the map-matching output points and the window distance.
In this way, we remove the error introduced by the vehicle placement in a location that does not entirely cover the real measurement.
At this point, the attack estimates the new coordinates for the vehicle after the time window.
To do so, we rely on a geodesic function (the closest line connecting two points in a curved surface) that computes new point coordinates, knowing the direction (heading) and the distance applied to the vehicle in the time $t_{window}$.
Finally, after the $max\_interpolaton\_points$ windows interpolation, the algorithm proceeds with the map-matching using the coordinates inferred by the physical model. 

At every interaction of the interpolation, the returned points from the map-matching are merged with the previously computed ones to form the final path. 
The operation repeats until there are no more packets to process.
The output of the dynamic map-matching algorithm is the inferred path starting from only the \ac{swa} sensor and speed data.

\section{Implementation}\label{sec:implementation}

A practical implementation of the attack requires the adoption of a low-resource hardware device.
This tool can be, for example, a USB2CAN device from \texttt{8 devices} (\url{https://shop.8devices.com}).
It can connect to the \ac{obd} port and allows the \ac{can}-Bus connection with a Linux system.
In our experiments, as discussed in the later section, we connected directly to our laptop, but a Raspberry or similar device can also be used.

To perform the attack, we implemented four auxiliary \textit{Python} scripts.
The first two files, namely \texttt{reWheel.py} and \texttt{AngleChecker.py}, assist the attacker during the manual reverse engineering phase in which the \ac{swa} must be found. The \texttt{OBDrop.py} script performs the logging phase on-board the vehicle, while \texttt{OPD-II.py} is the core of the attack, implementing the dynamic map-matching algorithm and outputs the inferred path. In the following, we describe the scripts in detail.
We start the attack description with the discussion of the procedure to choose the setting parameters.
The attack comprises six steps using these tools; Figure~\ref{fig:workflow} shows the attack's overall workflow and the result assessment.
Lastly, we add a further step for comparing the \ac{gpx} baseline for the path to provide an evaluation. \\

\noindent{\textbf{Tuning the parameters.}}
In the attacking tools, we set four custom parameters, namely, the time window length in seconds $t_{window}$, the minimum speed at which the car is considered in a straight trajectory $speed_{min}$, the maximum physical angle of the wheels $steer_{max}$, and the maximum number of points to use for calling the map-matching algorithm $max\_interpolation\_points$.
We performed a grid search with different values for these parameters:
\begin{itemize}
    \item $t_{window}=[0.05, 0.1, 0.5, 1]$;
    \item $speed_{min}=[40, 50, 60, 70, 80]$; 
    \item $steer_{max}=[30,35,40,45,50]$;
    \item $max\_interpolation\_points=[10,20,30,40,50]$.
\end{itemize}

\begin{figure}[!h]
    \begin{center}
    \includegraphics[width=.9\columnwidth]{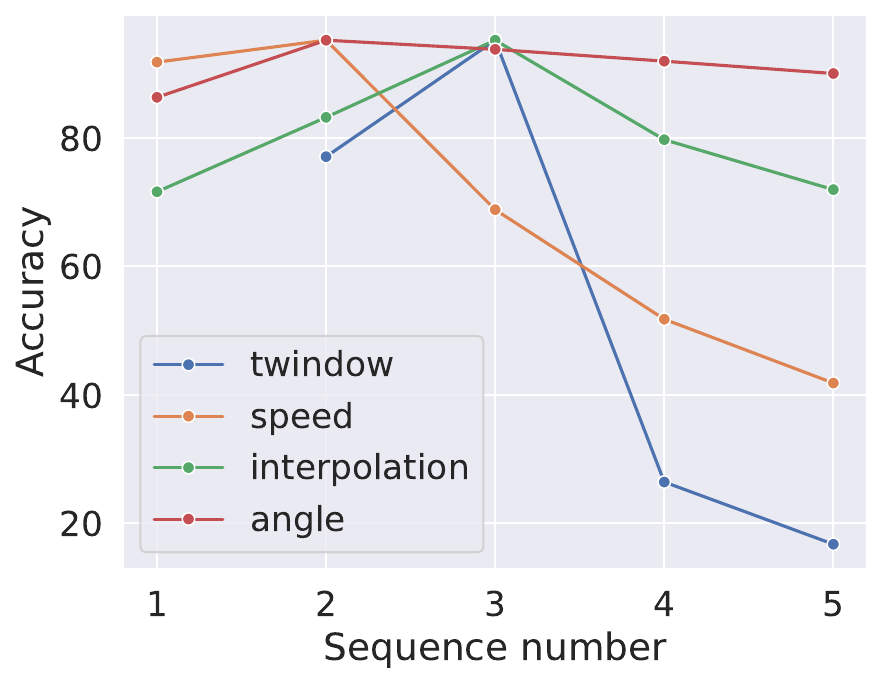}
    \end{center}
    \caption{\label{fig:fine_tuning} The impact of a parameter on the accuracy, fixing the other with the best value found through the grid search. The $t_{window}$ and $speed$ have the greater impact.}
\end{figure}

\noindent
We adopted the combination with the higher average accuracy: $t_{window} = 0.1$, $speed_{max}=50$, $steer_{max}=35$, and for the interpolation we have $max\_interpolation\_points=30$.
Figure~\ref{fig:fine_tuning} represents the average accuracy while fixing the best value for all the parameters except one to study the impact of such a variable.  
The sequence number represents the index on the list of values for that parameter, as reported in the bullet list.

We notice that variations on $steer_{max}$ have smaller effects on the accuracy of the path inference algorithm compared to other variables. This may be due to the fact that cars typically have similar capabilities in terms of steering, and the minor effect of this variable is compensated by the map-matching algorithm. On the other hand, the size of the time window highly impacts the accuracy. In particular, we notice that a larger time window increases the level of approximation of the traveled path, thus leading to decreased accuracy.

\begin{figure}
    \begin{center}
    \includegraphics[width=\columnwidth]{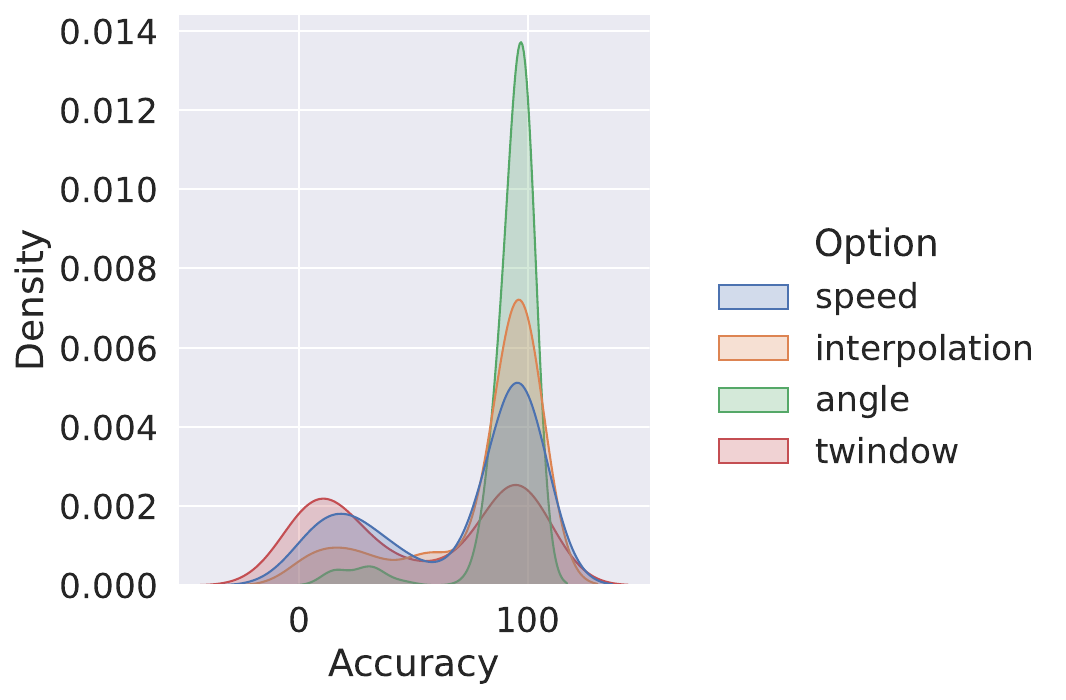}
    \end{center}
    \caption{\label{fig:fine_tuning_all} The distribution of the accuracy fixing three parameters at the ideal value while varying the fourth. The angle and the interpolation have a greater concentration of points around 90\%, while the time window variable negatively impacts the overall accuracy.}
\end{figure}

In Figure~\ref{fig:fine_tuning_all}, we report the accuracy in the same settings with every track used in the experiments for a better understanding of the influence of a parameter.
For each track, we compute the accuracy for every parameter using each value from the above lists.
As highlighted by the curves for the accuracy distribution, the $t_{window}$ and the $speed_{max}$ are the parameters that most influence the final results.
This is also visible in Figure~\ref{fig:fine_tuning}, where the accuracy for the time window and the speed drops with different parameters than the one found in the grid search.
Moreover, from the two figures, we can infer that the physical angle does not strongly impact the final results, consistently achieving an average accuracy of over 80\%.
The same discussion applies to the interpolation parameters, which have a slightly more significant impact than the physical angle.
In Appendix~\ref{appendix:fine_tuning}, we report the average accuracy for each setting in the grid search and the accuracy distribution for each parameter on each track varying the value adopted for the computation.
\\

\noindent{\textbf{SWA reverse engineering.}} 
The \texttt{reWheel.py} tool receives as input the log file generated by the attacker during the first car access, i.e., when they collect the steering wheel logs.
The tool returns the most probable IDs for the \ac{swa} sensor, along with the \textit{hamming distance} values for each byte in the data field to determine the ones containing the encoded data.
The tool allows the attacker to easily and immediately identify the most probable ID.
In addition, the script applies a filtering strategy to cut out lower IDs, specifically those lower than a threshold empirically set to \textit{0x300}. 
This value represents an approximation of the median ID value for the priority on the \ac{can}-Bus. 
This is due to the use of low IDs to broadcast the \ac{swa} information. 
Moreover, we found that some sensors repeat the same data in lower-priority IDs. 
We do not cover it, but we believe it is helpful for redundancy or other statistics in the vehicle.

The second tool, \texttt{AngleChecker.py}, supports the attacker during the identification of the decoding formula of the data field into angles.
The script allows the easy implementation, manipulation, and verification of the decoding for the data bytes.
We use equation \eqref{eq_angle}, using $c=0.01$ as a constant value to re-scale the angle sensor data after decoding it.
We set the value of $c$ empirically.
Since it is verified for all our four reverse-engineered vehicles, we can safely assume it is the same for the majority of the vendors.
Additionally, it already includes the decoding formula for the vehicles used during this work.
In this way, by combining the two tools, the attacker can reverse engineer the actual ID and bytes of the sensors in a semi-automatic manner.
All of these actions can be taken offline by the attacker.
In our GitHub repository, we include a sheet containing the already reverse-engineered sensors for specific vehicle models. \\

\begin{figure}[]
    \begin{center}
    \includegraphics[width=.8\columnwidth]{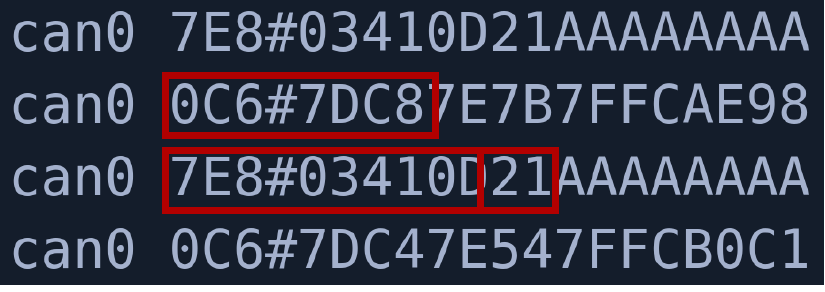}
    \end{center}
    \caption{\label{fig:packets}CAN messages containing the ID spreading information about the steering wheel angle sensor (upper red square) and the OBD answer with speed (lower red square).}
\end{figure}

\noindent{\textbf{Data collection.}} 
This step concerns the data logging and message injection in the \ac{can}-Bus.
The attacking device onboard the vehicle can run the \textit{OBDrop} tool (\texttt{OBDrop.py} script in the suite).
It is a simple yet very effective code utilizing \texttt{can-utils}, the Linux utilities to interact with a \ac{can} network. 
The script takes the \ac{swa} ID as an argument and then uses it to collect packets reporting the angle information.
At the same time, it continuously sends \ac{obd} requests to retrieve the vehicle's speed through the dedicated response ID \textit{0x7E8}, as discussed in Section~\ref{subsec:obdrop}.
The command is as follows:
\begin{verbatim}
Popen(['candump', '-t', 'A', '-l', 
f'{can_iface},7E8:7FF,{angle_id}:7FF'])
\end{verbatim}
\noindent
The $-t$ flag and parameter $A$ indicate the type of log we want to collect. 
In this case, it includes the entire packet's content and the date. 
The $-l$ flag tells \texttt{candump}, a tool from \texttt{can-util} suite for logging, to collect the packets in a file with extension \texttt{.log}. 
The last part of the command specifies the filter to apply during data collection and the interface in which it should be applied.
Except for the \ac{swa} ID, the filter clears the packets to get only the \ac{obd} responses.
The filter necessitates the mask to run, which is constant thanks to the type of IDs we are looking at (e.g., \textit{0x7FF} with no extended identifiers).  
As soon as the log starts, \texttt{OBDrop.py} sends packets using \texttt{cansend} utility specifying the \ac{obd} message asking for the speed, which is standardized across the vendors as we reported before:
\begin{verbatim}
7DF#02010DAAAAAAAAAA
\end{verbatim} 
As we report in Figure~\ref{fig:packets}, the $candump$ utility saves only the packets filtered out by the \ac{swa} ID, $0x0C6$ in the Figure, and the \ac{obd} response. 
The angle comprises the first two data bytes of the message that, in the experiment used for the figure, has an offset of $0x7FFF$ and a value of $0x7DC8$, that converted is $-5.67$, which indicates the wheels are slightly turned towards the right direction. 
Instead, the speed is easy to retrieve thanks to the standard. 
It is the fourth byte of the data section and is simply the velocity's hexadecimal value.
It has a minimum value of 0 and a maximum value of 255, as allowed using one byte, and the unit of measure is km/h. 
The Figure highlights the $0x21$ data, corresponding to 33 km/h.

At the end of the data collection, the attacker can pull the \texttt{.log} file back from the device, also using the Wi-Fi direct technology or an LTE connection while using a Raspberry Pi board, for example. \\

\noindent{\textbf{Path inference.}}
The \texttt{\ac{opd}} tool reconstructs the path traveled by the car. 
In particular, it uses the physical model of the vehicle over a time window that examines the data and advances in the route with a predefined time step.
This approach requires input information about the vehicle as we discussed in Section~\ref{subsec:opd}.
Once the program has the information about the initial point, the tool extracts the packets in a time window $t_{window}$ of 0.1 seconds.
The $t_{window}$ value is empirically tested, and it is the best solution to keep the physical model and movement of the cars as close to real as possible without losing algorithm accuracy.
During each time window, the distance computation uses the geodesic function provided by \texttt{Geopy Python library}~\cite{geopy}. 

We set the speed threshold $speed_{max}$ to 50 km/h to further correct the heading direction.
We assume a vehicle going straight cannot turn at a higher speed, or else it would have an accident.
We set that threshold to 50 km/h, which empirically shows the best performances for the algorithm.
In that case, the algorithm adjusts the bearing to the one between the starting and closing points of the previous time window, in which the car is taking the straight direction.
Also, the $steer_{max}$ value is set to 35\degree for the overall vehicle movement during the turning maneuver.
The condition is set by a trade-off between the accuracy of the algorithm and generalization for different vehicle models.

Following the algorithm, we implemented the map-matching tool using the service provided by \texttt{Valhalla}~\cite{valhalla} and the corresponding \texttt{Docker} container~\cite{valhalla_docker}, which makes it easier to use.
Valhalla uses the \texttt{Meili} utility, which includes routing and map-matching algorithms, such as the Viterbi algorithm and Hidden Markov Model, to retrieve the path along a road. 
The setup requires building a road graph provided by \texttt{Geofabrik}~\cite{geofabrick} and \texttt{OpenStreetMap}~\cite{osm}. 
Once the service is ready, the script uses \texttt{HTTP} requests to the \texttt{Docker} to get the points provided in \texttt{Meili}'s response.
For more details about \texttt{Valhalla} and \texttt{Meili}, we refer the reader to the \texttt{Valhalla} documentation~\cite{valhalla}.
All the matched coordinates are saved in a list that prints the map of the inferred path at the end of the process (when all packets are inspected) and saves them in a \texttt{\ac{gpx}} file. 
The $max\_interpolation\_points$ value for window interpolation, in our experiment set to 30, is also determined by the best accuracy possible.
It allows the physical model to perform a curve before the map-matching is applied to a smaller or greater system of points. 
In both cases, it leads to matching and distance error propagation.
If other packets are available, the process starts again with the new time window.
Ultimately, the script outputs the inferred path in a web window using the \texttt{Folium} python tool~\cite{folium}, highlighting the path on a \texttt{\ac{osm}}interactive map.
\\

\noindent{\textbf{Ground truth comparison.}}
We use this step only to evaluate the dynamic map-matching algorithm and attack.
While collecting the trips, we recorded the coordinates using a smartphone.
In the \texttt{OPD-II} tool, we first get the real path, aligning it with the same map-matching method previously described with Valhalla.
We only need to do the \texttt{Meili} request with the coordinates saved by the smartphone.
After that, we call the \texttt{cmpgpx}~\cite{cmpgpx} tool that applies the Needleman-Wunsch bioinformatics algorithm to align the points of the two sequences and compare them.
The tool's output is the overall similarity score used for the evaluation.
We describe the evaluation in detail in the next section.

\section{Evaluation}\label{sec:evaluation}
In this section, we discuss the experiments and evaluate this attack.
In Section~\ref{subsec:exp}, we describe how we perform the experiments and the paths, while in Section~\ref{subsec:res} we report the scores and the accuracy for the inferred routes against the baseline.

\subsection{Experiments}\label{subsec:exp}
The method comprehends the recording of the packets going through the car network while registering the \ac{gps} location during the trip.
This mode allows us to compare the crossed route with the inferred path in the output of our attack.
The configuration involves a laptop with Intel i7 and 16 Gigabytes of memory.
The memory is enough to run the setup for Valhalla using a big enough region from \texttt{Geofabrik} to build the graph.
Depending on the region, more memory could be needed.

The data collection requires the laptop to be connected to the \ac{obd} port.
In this matter, we use the \ac{obd}2 version of the Korlan UBS2CAN device from \texttt{8 devices} (\url{https://shop.8devices.com}). 
The USB2CAN integrates a \ac{can} adapter and converts the data to be used by libraries such as \texttt{can-utils} in Linux systems.

\begin{figure}
    \begin{center}
    \includegraphics[width=0.9\columnwidth]{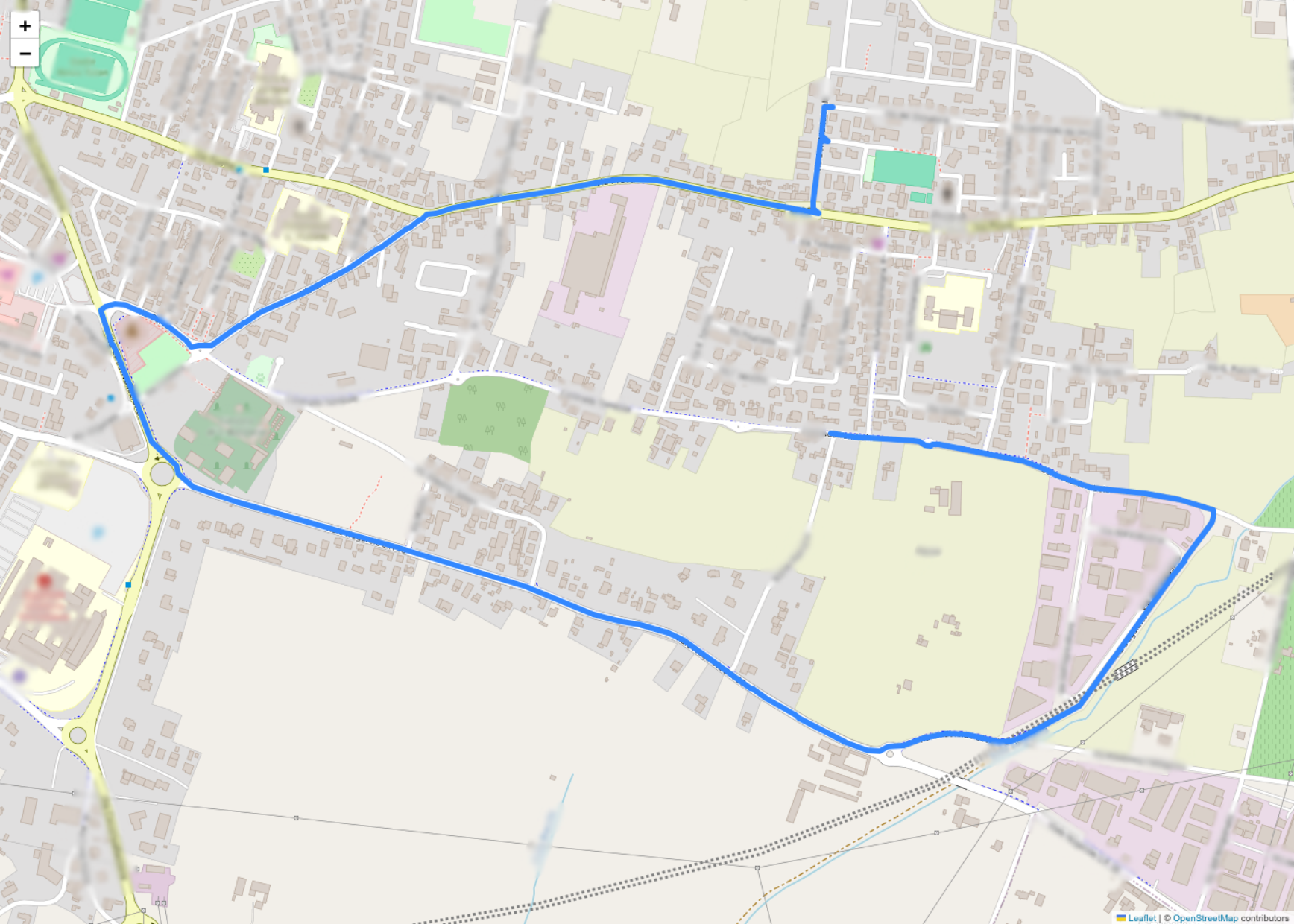}
    \end{center}
    \caption{\label{fig:opd_res} Result of the path inference showing the road crossed by the vehicle.}
\end{figure}

Initially, we register the vehicle's initial position and the direction it is pointing to.
During a trip, we use \textit{OBDrop} tool to collect the packets of interest (speed and \ac{swa} sensor).
At the same time, we register the path using a mobile phone and an app tracking the device's position.
In our case, we use the Outdooractive Android application (\url{https://www.outdooractive.com/}).
This app is mainly for outdoor activities such as hiking, but the tracking functionality works well even with a car.
The app also has the convenient \ac{gpx} file exporting function, so we have the ground truth path ready to use.
After that, we run the \textit{OPD-II} tool to infer the vehicle's path as described in the previous section.
Figure~\ref{fig:opd_res} represents an example of an inferred path and its visualization on \texttt{OpenStreetMap} using a browser and the \texttt{Folium} Python library~\cite{folium}.
We recorded the traces in different road scenarios and traffic conditions: city center, country roads, highways, and small towns. The road curvature and the speed are unique during each itinerary. The traffic difference comes mainly from the time of the trip, but it does not affect the path inference attack. 
In Figure~\ref{fig:opd_res}, we removed the street names in order to avoid possible trivial information leakage.
In Table~\ref{tab:experiments}, we summarize the experiments, defining them by vehicle type, total of kilometers traveled, and minimum and maximum distance for the paths with a vehicle model.
The traffic scenario is irrelevant in our experiments because the environment does not influence the inference. 
Using speed and angle values with dynamic map-matching allows us to avoid accuracy loss due to roundabouts, u-turns, stoplights, or other abnormal traffic situations. 
In the experimentation, we used four vehicles for a total of 41 trips and a length of 221.29 kilometers.
Table~\ref{tab:experiments} contains the information of each vehicle, such as production year and \ac{swa} sensor ID. 
\begin{table}[!h]
    \centering
    \caption{Experiments table}
    \resizebox{\columnwidth}!{
    \begin{tabular}{l|l|l|l|l|l}
    \hline
        \textbf{Model} & \textbf{Year} & \textbf{\ac{swa} ID} & \textbf{\# Tracks} & \textbf{Total km} & \textbf{Total accuracy (\%)} \\
    \hline
       \textbf{Renault Capture} & 2014 & 0x0C6 & 13 & 50.19 & 93.99  \\
       \textbf{Dacia Duster} & 2019 & 0x0C6  & 9 & 69.60 & 96.16 \\
       \textbf{Opel Crossland} & 2017 & 0x2F5 & 10 & 39.90 & 98.88\\
       \textbf{Peugeout 5008} & 2019 & 0x2EB & 9 & 61.60 & 94.35 \\
       \hline
       \multicolumn{3}{c|}{\textbf{Total}}  & \textbf{41} & \textbf{221.29} & \textbf{95.25} \\
    \hline
    \end{tabular}
    }
    \label{tab:experiments}
\end{table}

The result of the inference is then compared with the ground truth \ac{gps} signals collected during the trips, also processed through Valhalla and Meili.
To accomplish this, we use the tool introduced in the previous section: \textit{Cmpgpx}~\cite{cmpgpx}.
The juxtaposition of the two traces through the algorithm Needleman-Wunsch results in an overlapping score.

\begin{figure}
    \begin{center}
    \includegraphics[width=\columnwidth]{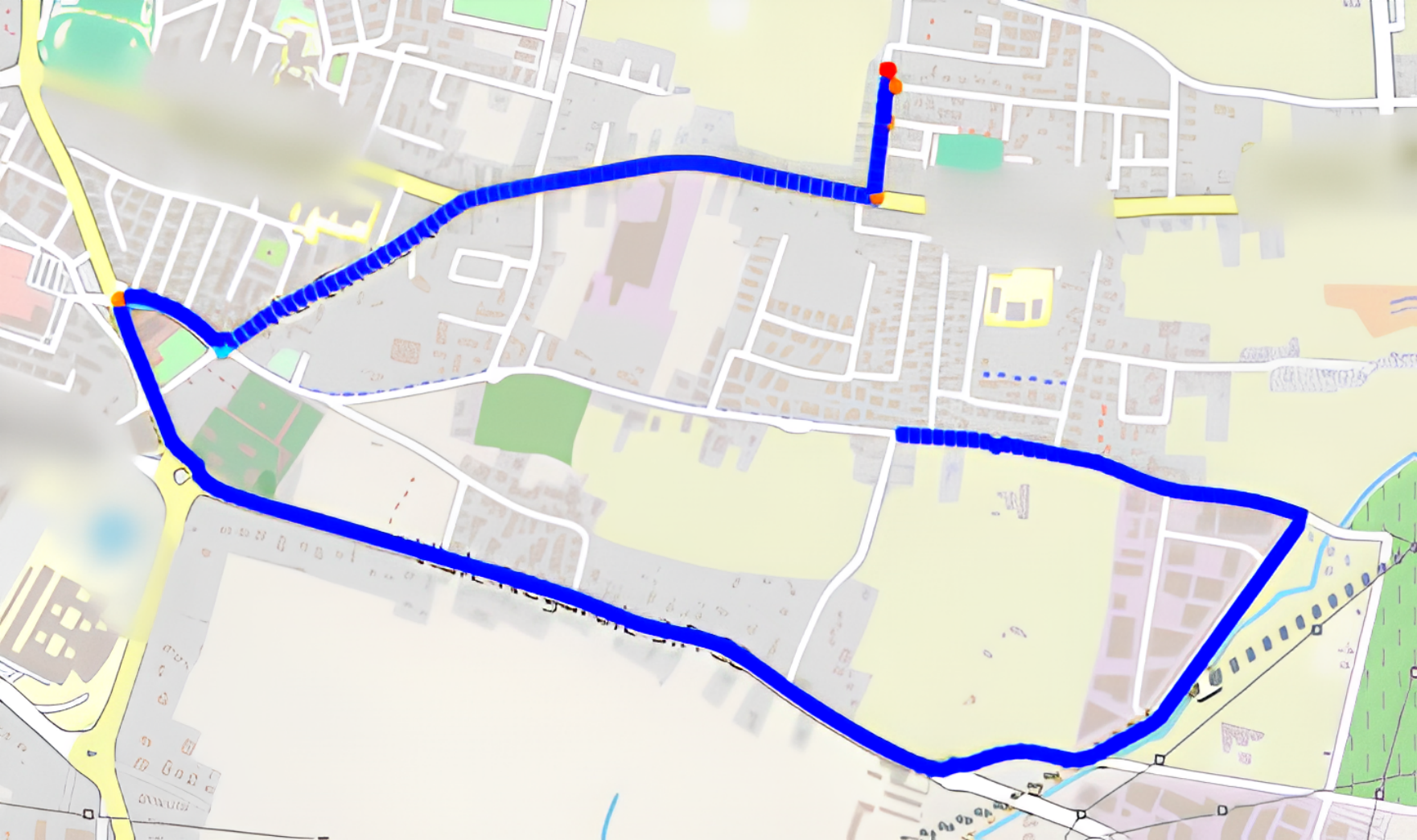}
    \end{center}
    \caption{\label{fig:path_plot} The comparison process graphically outputs the overlap between the ground truth GPX file and the inferred one. The red and orange dots are the parts missing in the inference.}
\end{figure}

We want to point out that we don't provide actual paths and positions in the results for anonymity.
The location of the experiments, all performed by the authors, could pinpoint the position and leak private information, harming the blind peer-review process.
Nonetheless, we provide an example of the output of \textit{Cmpgpx} in Figure~\ref{fig:path_plot}, always removing the street labels.
The Figure shows the common path between the traces, highlighting in red the few parts that are not overlapping.

\subsection{Results}\label{subsec:res}
Our experiment provides an overall accuracy of 95.25\%, with a minimum of 83.13\% and a maximum score of perfect reconstruction of 100\%.
The results outperform the state-of-the-art works, which in the best case scenario reaches, for~\cite{sarker}, on average, 89\% of accuracy to have the correct path in the top 5 tracks recovered. 
Instead, \cite{waltereit} obtains a 96.5\% path retrieval, sacrificing accuracy and dropping it to 75\%.
Moreover, the result of the path inference is deterministically determined by the starting point and the vehicle's movement inferred by the speed and the \ac{swa} sensor.
The algorithm outputs the track directly with high confidence without picking a rank for different possible results.
Sometimes, the baseline registration reduces the level of accuracy of the trip due to lower precision of the \ac{gps} signal on the mobile device.
Nonetheless, the results are significantly promising for this approach.

In Table~\ref{tab:experiments}, we report the total number of paths, kilometers, and overall accuracy for each vehicle tested.
We divided the data for the four models for which we carried out the reverse engineering part of the attack and covered various segments of different lengths.

\begin{figure}
    \begin{center}
    \includegraphics[width=\columnwidth]{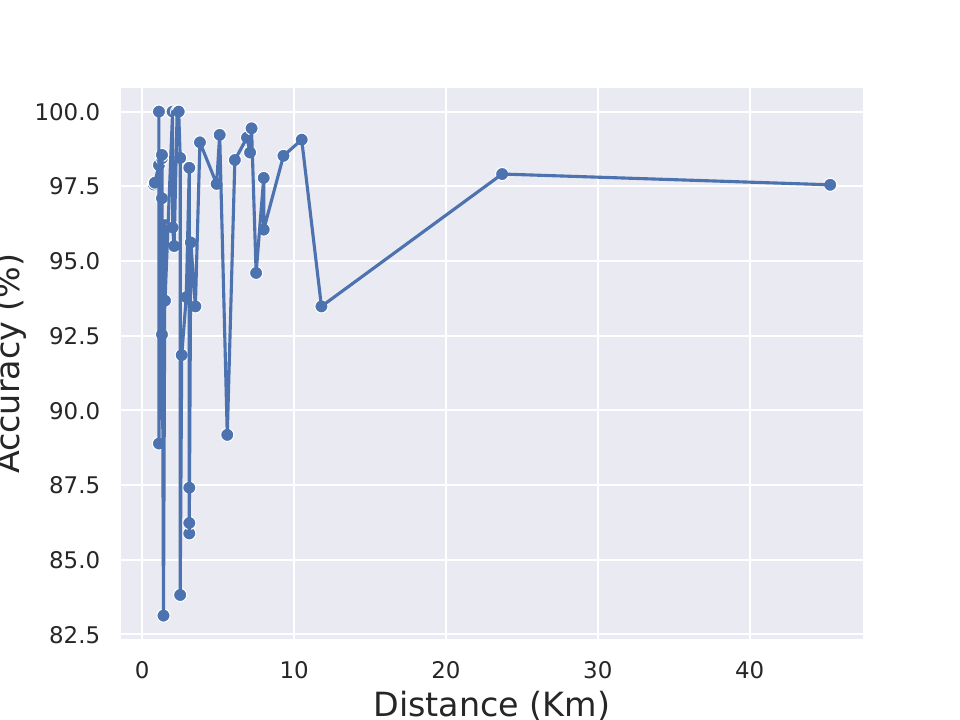}
    \end{center}
    \caption{\label{fig:accuracy} The accuracy in function of the path length. The result doesn't depend on the distance during the trip.}
\end{figure}

Additionally, in Figure~\ref{fig:accuracy}, we show the accuracy as a function of the track length.
As the plot shows, the accuracy doesn't depend on the distance but on the complexity of the road.
Especially, it is difficult to take the correct road in the presence of a dense cluster of points, as we discuss in Section~\ref{sec:discussion}.
Multiple roads very close to each other can trick the map-matching algorithm into positioning the vehicle on the wrong street, making it harder to return to the correct road.

\begin{figure}
    \begin{center}
    \includegraphics[width=\columnwidth]{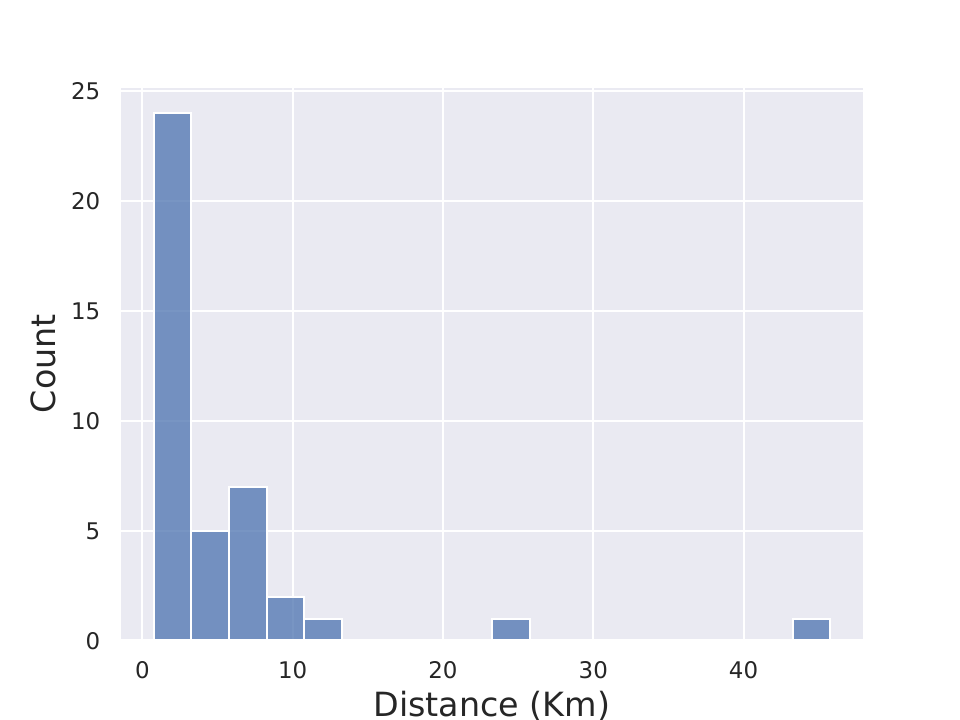}
    \end{center}
    \caption{\label{fig:count} Number of segments by the length. The majority of the trips fill within 1 km and 10 km. This is the average distance to reach different destinations, as well as those located in the same town.}
\end{figure}

Figure~\ref{fig:count} represents the number of tracks with a given size in an interval of 2.5 kilometers. 
Most of the journeys are between 0 and 10 kilometers, the general distance between towns in the area where tests were made. 
The longer path of 45.3 kilometers is the test performed on a highway, showing the possibility of tracking the vehicle even on long-distance trips.

\section{Discussion}\label{sec:discussion}
We propose an attack on the car's location privacy that doesn't rely on driving behavior and is not affected by external factors, such as traffic and speed limits.
The real-world application would be the position tracking of criminals performed by law enforcement when the lawbreakers know about the possibility of \ac{gps} tracking installation and the use of jammers.
The overall accuracy outperforms previous works on this topic.
Furthermore, the output is deterministic and doesn't need training in a particular area.
In this work, performances are not an issue due to using the Valhalla Docker container for the map-matching algorithm, which doesn't require many resources after initialization.
In fact, the \textit{OPD-II} tool doesn't build the road network graph itself, and this avoids the computation and memory overhead derived by the graph construction and operations.  

Nevertheless, this attack suffers from the presence of multiple interceptions in a small space, such as a roundabout crossed by a bridge or multiple intersections overlapped like in very complex junctions.
In these cases, different road options could have affected the algorithm, which could have taken the wrong path.
Another limitation is the need to know the initial position perfectly, especially the heading. 
An incorrect initial bearing can lead to the propagation of a heading error and, after multiple turns, lead the vehicle outside the correct path.
This is especially true if the vehicle goes on at a low speed, where the high-speed correction doesn't occur.
Despite that, the bearing rectification mainly adjusts this problem.

A defense proposal for this attack is out of the scope of this work because researchers have already presented different approaches in the literature.
One example can be to scramble the \ac{can} IDs after every timeout chosen in advance, as presented by Woo et al.~\cite{id_permutation}.
In this way, at a certain point that can also happen during the trip, the \ac{swa} sensor ID differs from the one found during the reverse engineering process, leading to a malfunction in the attack.
The problem of the car environment is the slow adoption and adaptation of these defenses and new technologies.

After the attack we present here, future works will focus on adopting a more complex and realistic physical model, for example, the Ackermann model. 
The more sophisticated the model, the more variables play in the definition of the vehicle.
This makes computations of the different physical aspects harder during the bearing update but can make it more precise. 
Another improvement is adjusting the algorithm to keep the vehicle on the road at each time window with less loss in precision on the path.
Also, the overall accuracy could benefit from the feedback on the possible inferred path from knowing intermediate points or the final destination if it differs from the starting point. 
Another potential improvement is using machine and deep learning models to predict the road and compare the path shape with the possible streets corresponding to it.
This approach would make the attack much more complex, while the simplicity of the approach we propose makes it available to every person with little knowledge of the \ac{can}-Bus and its components.

\section{Conclusion}\label{sec:conclusion}
The deterministic dynamic map-matching attack (\ac{opd}) and the tool suite we propose in this paper allow a malicious perpetrator or, in a benign scenario, the police investigation to resolve the path the driver pursues without following it.
The attack is straightforward and needs only readily available data on the \ac{can}-Bus and through the \ac{obd} protocol.
In fact, the use of the simple bicycle model allows the attacker to extract the path from the two already discussed information: speed and steering wheel angle value.
Combining the physical model and applying a map-matching algorithm step-by-step permits the reconstruction of the \ac{gps} locations crossed by the vehicle with high accuracy, that is, around 95\%.
The attack once again shows privacy problems in the automotive scenario and a delay in the manufacturer's application of defenses.
At the same, the attack can also be helpful to law enforcement corps during investigations and can be exploited until \ac{oem}s don't apply a real patch to this problem.

\bibliographystyle{ACM-Reference-Format}
\balance 
\bibliography{biblio.bib}

\appendix

\section{Bicycle Model}\label{appendix:bicycle}
\begin{figure}[!h]
    \begin{center}
    \includegraphics[width=\columnwidth]{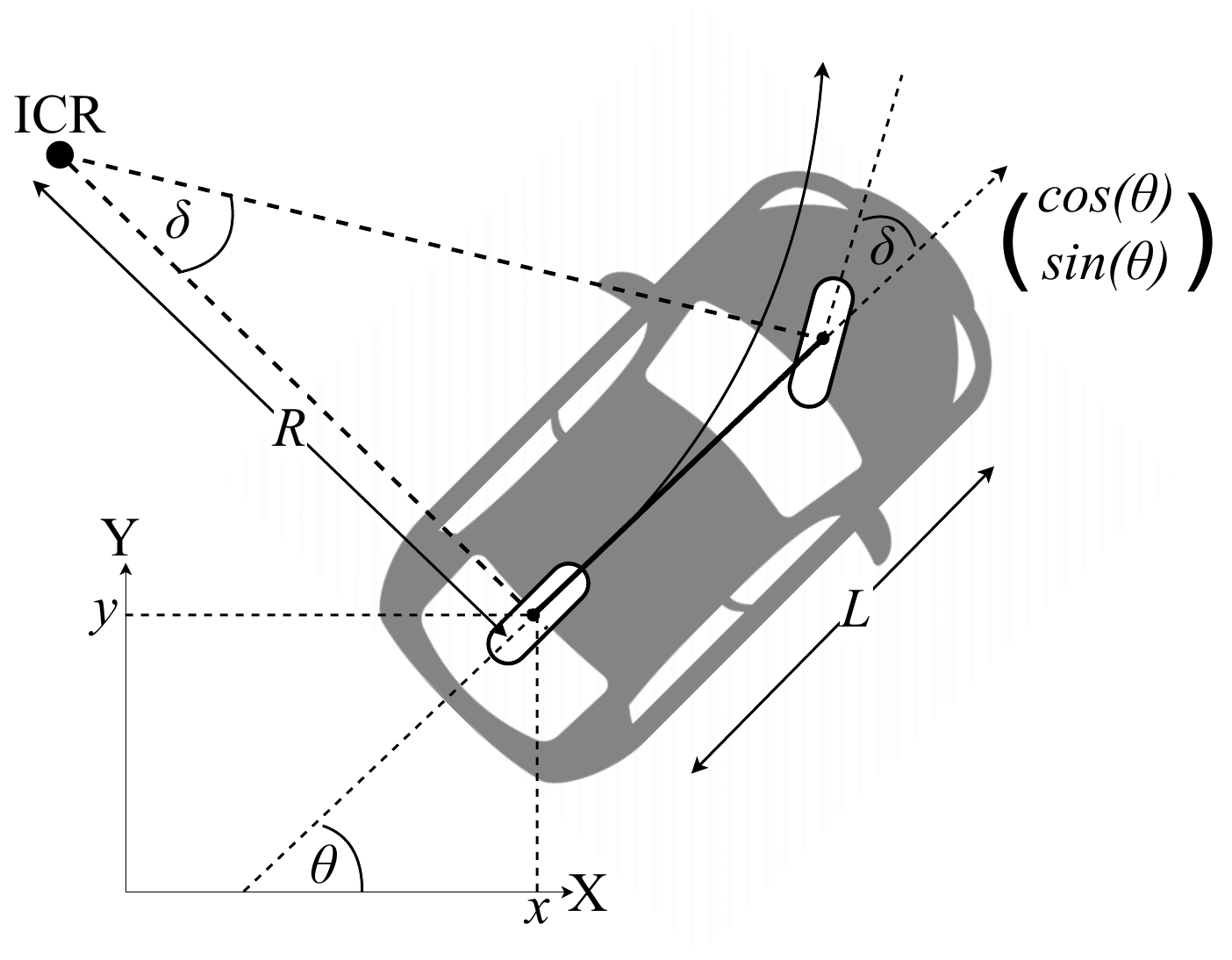}
    \end{center}
    \caption{\label{fig:bicycle} Bicycle model representation.}
\end{figure}

Here, we present the physical model adopted in this work.
This is the simplest model available for four-wheel-based vehicles.
It captures vehicle motion well in a typical driving environment but also presents some limitations. 
For example, it doesn't consider the lateral slip of the wheels.
Essentially, it simplifies the four-wheel into a two-wheel system.
Figure~\ref{fig:bicycle} represents the model and the motion computation.
The front wheel controls the vehicle's heading, and $\theta$ is the heading angle, while $\delta$ is the steering angle.

We can analyze it from a reference point: the center of the rear axle or wheel.
The \ac{icr} is the point at which the body has zero velocity in a particular instant of time.
We can compute the angular speed $\omega$ knowing \textit{R} and the speed \textit{v}, that is
\begin{equation}
    R = L / tan(\delta),
\end{equation}
\begin{equation}
    \omega = v / R = v \times tan(\delta) / L,
\end{equation}
\noindent
where \textit{L} is the wheelbase of the vehicle (distance between the front and rear axes).
Now, it is possible to compute the next state of the vehicle.
Simply, the new coordinates are:
\begin{equation}
    x_{new} = v \times cos(\omega),
\end{equation}
\begin{equation}
    y_{new} = v \times sin(\omega).
\end{equation}

The steering rate $\Phi$, that is, the rate of change of steering angle $\delta$, determines the heading change. 
Using a time window $t$ in which to compute the new heading leads to the formulation we provide in Equation~\ref{eq:new_angle}, here generalized:
\begin{equation}
        \delta = atan(sin(\Omega + \omega \times t), cos(\Omega + \omega \times t)).
\end{equation}

The kinematic bicycle model presents other equations depending on the reference point: the front axle or the center of gravity (middle of the wheelbase).
To understand our work, the lecturer only needs the computation starting from the rear axle as the reference point.
Additionally, we only account for the heading change due to using coordinates that use latitude, longitude, and bearing.

\section{Tuning the parameter}\label{appendix:fine_tuning}
\begin{figure*}[]
     \centering
     \begin{subfigure}{0.49\textwidth}
        \centering
        \includegraphics[width=0.75\columnwidth]{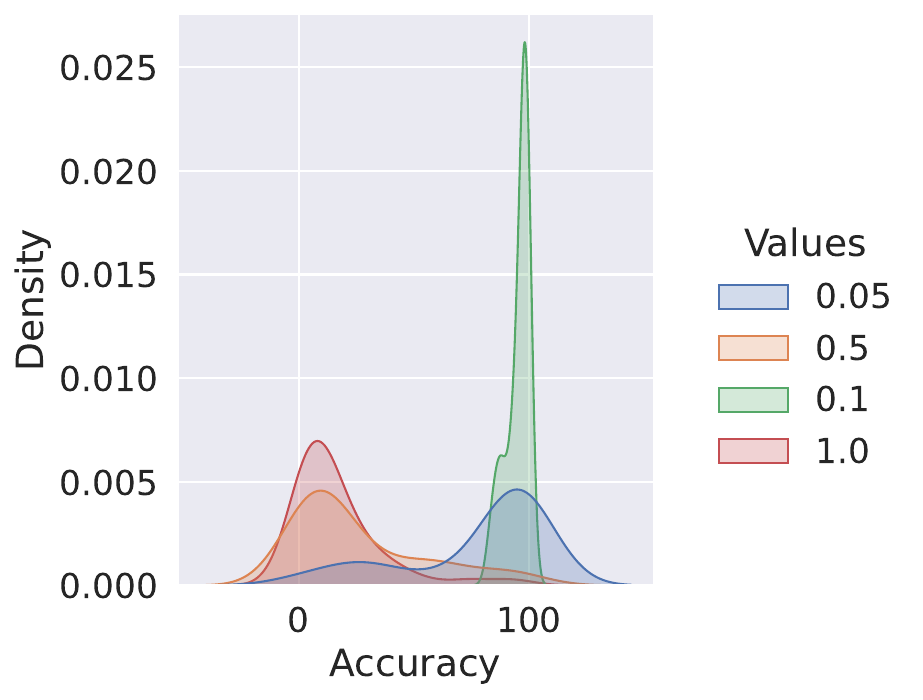}
        \caption{\label{fig:twin} Time window impact.}
     \end{subfigure}
     \hfill
     \begin{subfigure}{0.49\textwidth}
         \centering
             \includegraphics[width=0.75\columnwidth]{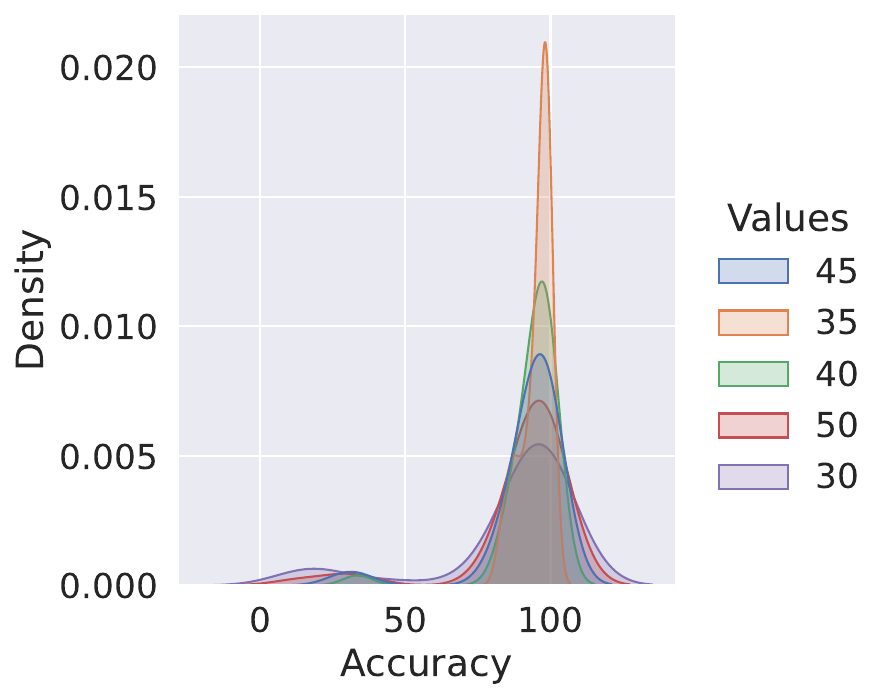}
            \caption{\label{fig:angle} Angle impact.}
     \end{subfigure}
     \hfill
     \begin{subfigure}{0.49\textwidth}
         \centering
         \includegraphics[width=0.75\columnwidth]{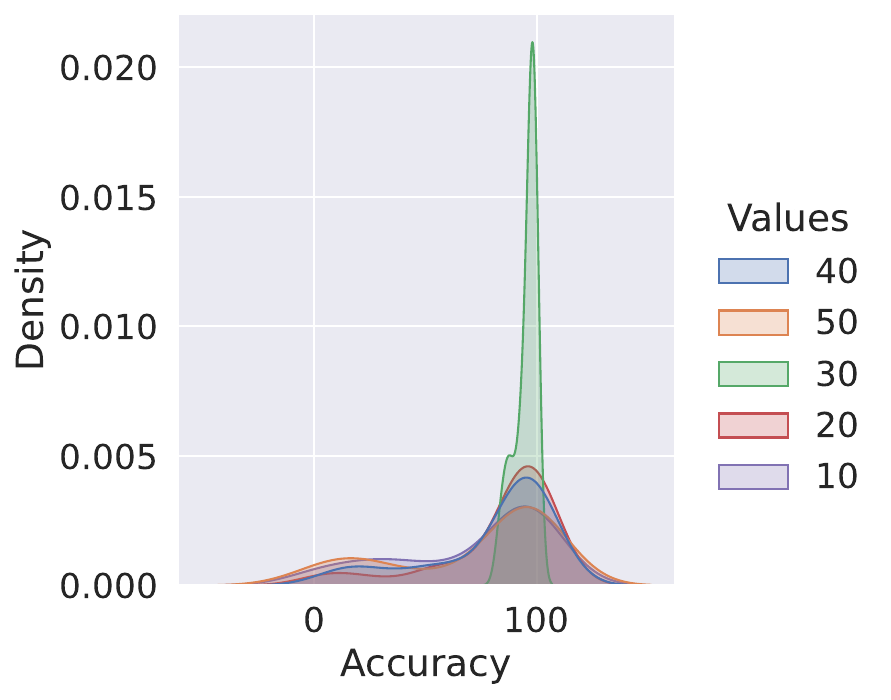}
        \caption{\label{fig:interpol} Interpolation impact.}
     \end{subfigure}
     \hfill
     \begin{subfigure}{0.49\textwidth}
         \centering
         \includegraphics[width=0.75\columnwidth]{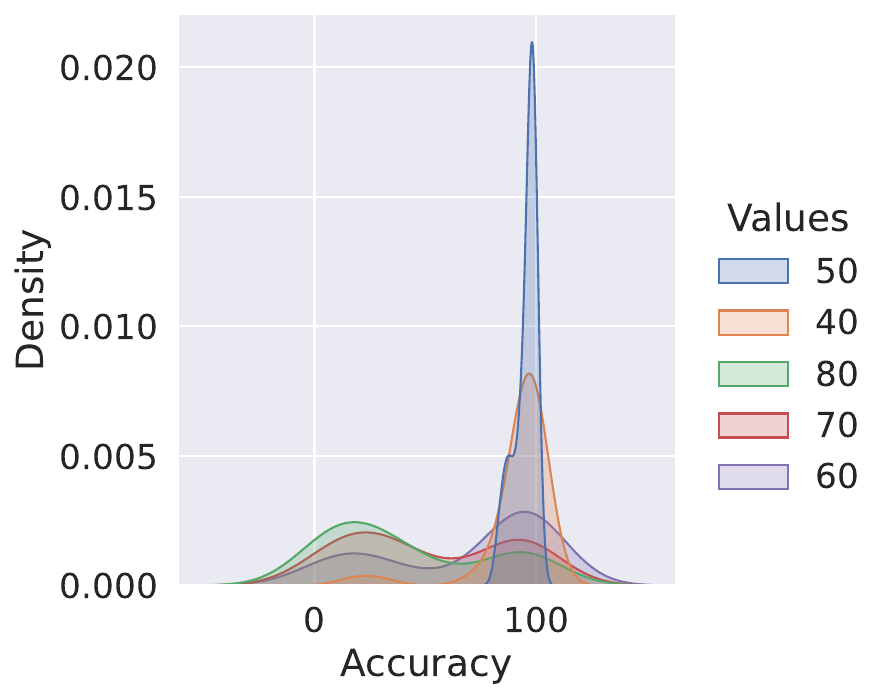}
        \caption{\label{fig:speed} Speed impact.}
     \end{subfigure}
        \caption{Impact of the different parameters}
        \label{fig:impact_params}
\end{figure*}

\begin{figure}[!h]
    \begin{center}
    \includegraphics[width=0.7\columnwidth]{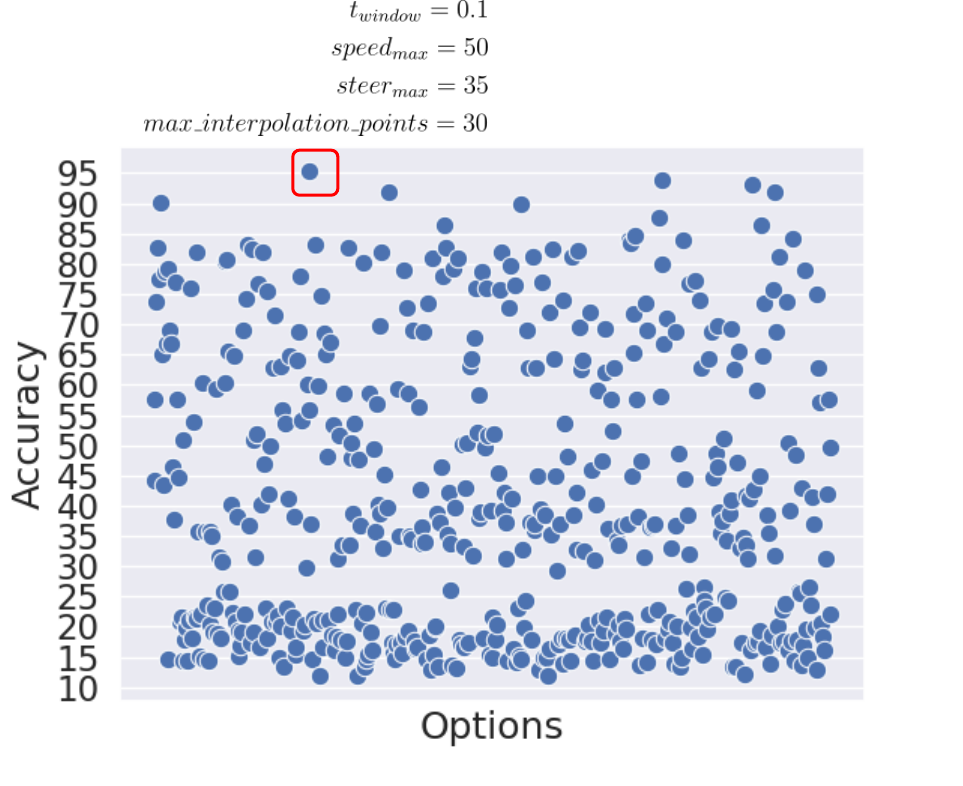}
    \end{center}
    \caption{\label{fig:avg_grid} Average accuracy of every combination of parameters using the grid search process. The highlighted result represents the combination we presented in Section~\ref{sec:implementation}.}
\end{figure}

The grid search aims to find the parameters combination that achieves the best scores on average on the tracks. 
In Figure~\ref{fig:impact_params}, we show the impact of every value in the parameter selection while keeping the other fixed with the optimal value found during the grid search. 
Figure~\ref{fig:avg_grid} represents the average accuracy for all 500 combinations. 
We highlight the best result that is the output of our algorithm presented in Section~\ref{sec:evaluation}.
As reported in Section~\ref{sec:implementation}, the significant impact is dictated by the time window and the maximum speed, but for every parameter, the change in the value can bring a considerable reduction in the accuracy in each track. 
In fact, we can see that the optimal value of 30 carries a significantly better accuracy distribution for the interpolation parameter.

\section{Online Resources}\label{appendix:resources}

We provide all the project's code on GitHub at this link: \url{https://anonymous.4open.science/r/OPD-II-9FCB}
The repository contains all the information about the necessary hardware and software. 
Each step of the experiment is reproducible by following the description and instructions.
We provide a \textit{csv} file containing the accuracy and length of each segment of the experiments.
We also want to stress that we don't provide examples and experiment segments at the moment of submission for privacy reasons and to avoid the possibility of inferring the authorship of this work.

\end{document}